\documentclass[twocolumn,twocolappendix,trackchanges]{aastex631}

\newcommand{\pnt}[1]{{\scriptstyle#1}}
\usepackage{hyperref}
\usepackage{roboto}
\usepackage{threeparttable}
\usepackage{amsmath}
\usepackage{enumitem}
\usepackage{orcidlink}

\begin{document}

\title{Towards unveiling the Cosmic Reionization: the ionizing photon production efficiency ($\xi_{ion}$) of Low-mass H$\alpha$ emitters at $z \sim 2.3$}

\author{Nuo Chen \orcidlink{0000-0002-0486-5242}}
\affiliation{Department of Astronomy, Graduate School of Science, The University of Tokyo, 7-3-1 Hongo, Bunkyo-ku,
Tokyo 113-0033, Japan; \url{nuo.chen@grad.nao.ac.jp}}
\affiliation{National Astronomical Observatory of Japan, 2-21-1 Osawa, Mitaka, Tokyo 181-8588, Japan} 

\author{Kentaro Motohara \orcidlink{0000-0002-0724-9146}}
\affiliation{Department of Astronomy, Graduate School of Science, The University of Tokyo, 7-3-1 Hongo, Bunkyo-ku,
Tokyo 113-0033, Japan; \url{nuo.chen@grad.nao.ac.jp}}
\affiliation{National Astronomical Observatory of Japan, 2-21-1 Osawa, Mitaka, Tokyo 181-8588, Japan}
\affiliation{Department of Astronomical Science, SOKENDAI, 2-21-1 Osawa, Mitaka, Tokyo 181-8588, Japan}

\author{Lee Spitler \orcidlink{0000-0001-5185-9876}}
\affiliation{School of Mathematical and Physical Sciences, Macquarie University, Sydney, NSW 2109, Australia}
\affiliation{Macquarie University Astrophysics and Space Technologies Research Centre,  Macquarie University, Sydney, NSW 2109, Australia}
\affiliation{Australian Astronomical Optics, Faculty of Science and Engineering, Macquarie University, Macquarie Park, NSW 2113, Australia}

\author{Kimihiko Nakajima \orcidlink{0000-0003-2965-5070}}
\affiliation{National Astronomical Observatory of Japan, 2-21-1 Osawa, Mitaka, Tokyo 181-8588, Japan}

\author{Yasunori Terao}
\affiliation{Institute of Astronomy, Graduate School of Science, The University of Tokyo, 2-21-1 Osawa, Mitaka, Tokyo 181-0015, Japan}

\begin{abstract}
We investigate the galaxy properties of $\sim$400 low-mass ($<10^9\,M_{\odot}$) H$\alpha$ emitters (HAEs) at z $\sim$ 2.3 in the ZFOURGE survey. The selection of these HAEs is based on the excess in the observed $K_s$ broad-band flux compared to the stellar continuum estimated from the best-fit SED. These low-mass HAEs have elevated SFR(H$\alpha$) above the star formation main sequence (SFMS), making them potential analogs of the galaxies that reionized the universe during the epoch of reionization. The ionizing photon production efficiencies ($\xi_{ion}$) of the low-mass HAEs have a median value of $\mathrm{log}(\xi_{ion}/erg^{-1} Hz)=25.24^{+0.10}_{-0.13}\  (25.35^{+0.12}_{-0.15})$, assuming the Calzetti (SMC) curve for the stellar continuum dust correction. This value is higher than that of main sequence galaxies by $\sim$0.2 dex at similar redshift, indicating that the low-mass HAEs are more efficient in producing ionizing photons. Our results also consolidate the trend of increasing $\xi_{ion}$ with redshift, but reveal a "downsizing" relationship between $\xi_{ion}$ and stellar mass ($M_{\odot}$) with increasing redshift. We further explore the dependence of $\xi_{ion}$ on other galaxy properties, such as the UV spectral slope ($\beta_{\mathrm{UV}}$), the UV magnitude ($M_{\mathrm{UV}}$), the equivalent widths ($EWs$) of H$\alpha$ and [O{\sc iii}] emission lines. Galaxies with the bluer UV slopes, fainter UV luminosities and higher equivalent widths exhibit elevated $\xi_{ion}$ by a factor of $\sim$2 compared to the median $\xi_{ion}$ of our sample. JWST data will provide an opportunity to extend our method and further investigate the properties of low-mass galaxies at high redshifts.
\end{abstract}

\keywords{galaxies: evolution – galaxies: high-redshift - galaxies: star-formation -  galaxies: dwarfs – cosmology: observations - reionization}
\let\clearpage\relax

\section{Introduction} 
\label{sec:intro}
The Epoch of Reionization (EoR) is a significant period in the cosmic history, when the ubiquitous neutral intergalactic medium (IGM) was ionized by the enormous amounts of ionizing photons (also called LyC or Lyman continuum) from the first luminous sources at that epoch. Current understanding suggests that reionization happens at $z > 6$ \citep[e.g.,][]{Pentericci14, Schenker14}. Several important aspects of the EoR still remain unclear, such as identifying the specific sources drive reionization \citep[]{Madau15, Robertson15}. Some studies propose that low-mass galaxies are expected to be major contributors to the cosmic reionization as the UV radiation budget is dominated by faint galaxies from the UV ($\sim1500\ \,\textrm{\AA}$) luminosity function at $z > 6$ \citep[e.g.,][]{Atek15, Bouwens15,Finkelstein19}. On the other hand, massive and bright galaxies are also proposed as major LyC leakers because of their larger escape fraction in model estimations \citep[e.g.,][]{Gnedin08, Naidu20}.

In order to fully understand how the universe was reionized, we must constrain the following, (i) the ionizing photon production efficiency, $\xi_{ion}$, defined by the number of Lyman continuum (LyC) photons, i.e, ionizing photons, produced per UV luminosity \citep[]{Robertson13} and (ii) the escape fraction of ionizing photons, $f_{esc}$, defined as a ratio of transmitted ionizing photons ($\lambda < 912 \mathrm{\AA}$) to input ionizing photons. Rest-frame optical emission lines serve as valuable tools for determining these parameters. The primary method for estimating $\xi_{ion}$ involves using nebular recombination lines, such as $\mathrm{H\alpha},$ to infer the ionizing photon production rate. $f_{esc}$, though derived from LyC flux emitted, exhibits correlations with various observational features of emission lines. However, accurately measuring these parameters at $z > 6$ presents considerable challenges. Directly measuring LyC flux is impossible due to the attenuation from the IGM along the line of sight. Meanwhile, galaxies at that epoch are too faint to construct a large and precise sample of emission line information.
 
Nevertheless, by conducting a detailed study of lower-redshift analogs, we can gain valuable insights into the physical properties of these early, faint galaxies at $z > 6$. Previous studies have already identified several low-redshift analogs. Individual galaxies with strong LyC leakage \citep[]{Izotov16a, Izotov16b, Izotov18a, Izotov18b} are characterized by high emission line ratios of $\mathrm{O32} = \mathrm{[O\pnt{III}]}\,/\,\mathrm{[O\pnt{II}]} \gtrsim 5$. This high O32 ratio has been suggested to be positively correlated with $f_{esc}$ \citep[]{Nakajima14,Izotov18a}. Also, the inter-dependence of strong $\mathrm{[O\pnt{III}]}$ emission line and $f_{esc}$ is pointed out by \citet[]{Nakajima14} through photoionization models. In another aspect, at lower redshifts, the estimation of $\xi_{ion}$ for individual galaxies at lower redshift has already been achieved using nebular recombination lines \citep[e.g.,][]{Bouwens16, Nakajima16, Shivaei18, Nanayakkara20, Emami20, Atek22, Stefanon22, Matthee22}.

Previous studies on $\xi_{ion}$ have used various samples: (i) Large sample of relatively massive galaxies ($>10^{9.5}\,M_{\odot}$) \citep[e.g.,][]{Shivaei18}, (ii) Small sample of low-mass galaxies \citep[e.g.,][]{Hayashi2016,Emami20}, (iii) Large sample of low-mass galaxies but at relatively low redshift \citep[$z \sim 1$; e.g.,][]{Atek22}. To gain a comprehensive understanding of $\xi_{ion}$ closer to EOR, pushing forward to large sample of low-mass galaxies at high redshift would be an important and necessary task.
A large number of low-mass H$\alpha$ emitters (HAEs) have been revealed at $z \sim 2.3$ by \citet[]{Terao2022}. These galaxies exhibit strong H$\alpha$ emission lines primarily produced by massive stars (O and B stars) with lifetimes of less than 10 Myr. The possible presence of high-ionization states in these galaxies makes it reasonable to assume these low-mass HAEs as analogs of LyC leakages during EoR.
Exploring the properties of them is therefore valuable in enhancing our understanding of cosmic reionization. In this work, we focus on a substantial sample of 401 low-mass H$\alpha$ emitters (HAEs) at $2.05 < z < 2.5$, which are identified using the photometric catalog from the FourStar galaxy evolution survey (ZFOURGE). The sample reaches down to a stellar mass of $10^8\,M_{\odot}$. For a comprehensive understanding of the selection criteria and the parent HAEs catalog, we refer readers to the detailed description provided in \citet{chen2023} (hereafter C24). 

The outline of this paper is as follows. We introduce the data in Section \hyperref[sec:obv]{2} and measurements of physical parameters in Section \hyperref[sec:samp]{3}. We exhibit the relationship between $\xi_{ion}$ and various galaxy properties of the selected low-mass HAEs in Section \hyperref[sec:pro]{4}. We further explore the observation and modeled results of $\xi_{ion}$ and discuss the implications of low-mass galaxies for their contribution to the cosmic reionization in Section \hyperref[sec:dis]{5}. Finally we summarize our result in Section \hyperref[sec:conclu]{6} and discuss future observations to further investigate the properties of HAEs at $z \sim 2.3$.

Throughout this thesis, we adopt the AB magnitude system \citep{Oke83}, assume a Chabrier(\citeyear{Chabrier03}) initial mass function (IMF) and a $\mathrm{\Lambda CDM}$ cosmology with $H_0 = \mathrm{70\ km\ s^{-1}\ Mpc^{-1}}$, $\Omega_m = 0.3$, and $\Omega_{\Lambda} = 0.7$.

\section{The galaxy sample}
\label{sec:obv}
In this study, we utilize the HAEs catalog from C24, which comprises 1318 HAEs at $\mathrm{2.05 < z < 2.5}$ from the ZFOURGE survey \citep{Straatman16}. These HAEs are obtained from the flux excess in the observed ZFOURGE-$K_s$
broad-band to the stellar continuum estimated from the best-fit SED. The catalog includes information such as redshift, observed emission line fluxes, and SED-derived properties (e.g., stellar mass, dust attenuation) for the HAEs, with IDs corresponding to the ZFOURGE catalog.
C24 refined the redshift information by incorporating spectroscopic redshifts ($z_{spec}$) from the MOSDEF survey \citep{Kriek15} and grism redshifts ($z_{grism}$) from the 3D-HST Emission-Line Catalogs \citep{Momcheva16}. The median redshift of the HAEs is at $\mathrm{z_{med} = 2.25}$.

C24 conducted rigorous SED fitting to obtain accurate galaxy properties by using the \robotoThin{CIGALE} code \citep{Boquien19}. The SED fitting, coupled with the reliable stellar continuum, allows for the derivation of emission line fluxes for H$\alpha$ and [O{\sc iii}] based on the flux excesses ($F_{excess}$) in the deep ZFOURGE near-IR broad- and medium-bandwidth filters. The derived emission line fluxes exhibit excellent agreement with spectroscopic and grism measurements from the MOSDEF and 3D-HST Emission-Line Catalogs. Rest-frame equivalent widths ($EWs$) of H$\alpha$ and [O{\sc iii}] emission lines are calculated from the $F_{excess}$ and the continuum flux density from the best-fit SED model. In Table \hyperref[tab:emission]{1}, we present the emission lines and the information of corresponding filters used in this analysis.

\begin{table}[t]
    \small
    \centering
    \label{tab:emission}
    \caption{The ZFOURGE data used to derive emission line fluxes}
    \begin{tabular}{ccc}
        \hline\hline
        Field & Emission line & Filter$\mathrm{\,^{a, b}}$  \\
         &  & (5$\sigma$ depth, 0.6")  \\
        \hline
        CDFS & $\mathrm{H\alpha}$  & $K_s\,(26.2)$  \\
        ($128\,\mathrm{arcmin}^2$)& $\mathrm{[O\pnt{III}]}$ & $H_s\,(24.9)$ or $H_l\,(25.0)$  \\
        \hline
        COSMOS & $\mathrm{H\alpha}$  & $K_s\,(25.5)$   \\
        ($135\,\mathrm{arcmin}^2$)& $\mathrm{[O\pnt{III}]}$ & $H_s\,(25.1)$ or $H_l\,(24.9)$  \\
        \hline
        UDS & $\mathrm{H\alpha}$ & $K_s\,(25.7)$  \\
        ($189\,\mathrm{arcmin}^2$)& $\mathrm{[O\pnt{III}]}$ & $H_s\,(25.1)$ or $H_l\,(25.2)$  \\     
        \hline
    \end{tabular}
    \begin{tablenotes}
    \item \textbf{Notes.} ${ }^{\text {a }}$ The H$\alpha$ emission line at $2.05 < z < 2.5$ redshifts into the $K_s$ filter. \\
    ${ }^{\text {b }}$ The [O{\sc iii}] emission lines redshift into $H_s$ filters at $z < 2.26$, and into $H_l$ filters at $z > 2.26$.\\
    \end{tablenotes}
\end{table}

\section{Measurements}
\label{sec:samp}
In this work, we quantify the star formation rates (SFRs) of galaxies by two indicators, $\mathrm{H\alpha}$ and FUV ($\mathrm{1500\AA}, L_{1500}$) luminosities. For the UV luminosity measurement, galaxies are required to be detected in ZFOURGE $B$ and $V$ filter, which yields a cut of 2.5\% of total HAEs. We compute the observed UV continuum ($L_{1500,uncor}$) and the UV slope ($\beta_{\mathrm{UV}}$) over a rest-frame wavelength range of $\mathrm{1400-2800\AA}$ by performing a multi-band fitting to broad-band photometry with the relation of $f_\lambda \propto \lambda^{\beta_{\mathrm{UV}}}$. For ZFOURGE-COSMOS field, the fitting includes broad-band photometry of $B, G, V, R, R_p, I, Z, Z_p$. For ZFOURGE-UDS and ZFOURGE-CDFS field, the fitting includes broad-band photometry of $B, V, R, I, Z$. The amount of dust attenuation is measured through the Bayesian SED fitting, providing the dust extinction $E(B-V)_{star}$, $E(B-V)_{neb}$. We parameterize
the difference between $E(B-V)_{star}$ and $E(B-V)_{neb}$ by a factor $f=0.8$ \citep{Saito20}.
Once we obtain the intrinsic UV luminosity ($L_{1500,cor}$) and intrinsic $\mathrm{H\alpha}$ luminosity ($L_{\mathrm{H\alpha},cor}$), the star formation rates are converted by using the calibration of \citet{Kennicutt12} with a correction to the \citet{Chabrier03} IMF.

The ionizing photon production efficiency $\left(\xi_{\text {ion}}\right)$ is defined as the ratio of the production rate of ionizing photons, $N\left(\mathrm{H}^0\right)$, in units of $\mathrm{s}^{-1}$ to the intrinsic UV luminosity $\left(L_{\mathrm{UV,cor}}\right)$ in units of $\mathrm{erg\,s^{-1}Hz^{-1}}$:
\begin{equation}
\label{equ:xiion}
\xi_{ion}=\frac{N\left(\mathrm{H}^0\right)}{L_{\mathrm{UV,cor}}}\left[\mathrm{s}^{-1} / \mathrm{erg}\,\mathrm{s}^{-1} \mathrm{~Hz}^{-1}\right].
\vspace{0.01cm}
\end{equation}
In this work, $L_{\mathrm{UV,cor}}$ is derived at the rest-frame $1500 \AA$, and $N\left(\mathrm{H}^0\right)$ is calculated as follows. For an ionization-bounded nebula, we could derive $N\left(\mathrm{H}^0\right)$ from the intrinsic $\mathrm{H\alpha}$ luminosity, $L_{\mathrm{H\alpha},cor}$, through the relation of \citet{Leitherer95}:
\begin{equation}
\label{equ:xiion2}
N\left(\mathrm{H}^0\right)\left[\mathrm{s}^{-1}\right]=\frac{1}{1.36} \times 10^{12}\,L_{\mathrm{H\alpha},cor}\left[\mathrm{erg}\,\mathrm{s}^{-1}\right].
\end{equation}
One important assumption in calculating $\xi_{ion}$ above is the zero LyC escape fraction, i.e, $f_{esc} = 0$, since we assume the rate of recombinations balances the production rate of ionizing photons. The combination of Equation \hyperref[equ:xiion]{1} and \hyperref[equ:xiion2]{2} actually gives us $\xi_{ion,0}$, while the intrinsic $\xi_{ion}$ should be $\xi_{ion} = \xi_{ion,0}/(1-f_{esc})$. 

It is estimated that if the universe is totally reionized by galaxies, a large $f_{esc} \sim 20\%$ at $z > 7$ is needed \citep{Ouchi09}. In contrast, after the universe is fully reionized, based on direct LyC imagings at lower redshift \citep[$2<z<4$, e.g.,][]{Vanzella10,Grazian16,Matthee17}, an upper limit of the escape fraction is $f_{esc} < 6\%$. Therefore, ignoring the mentioned $f_{esc}$ correction would not significantly alter the trends we find in this analysis. We will assume $\xi_{ion} \simeq \xi_{ion,0}$ when investigating the relationship between $\xi_{ion}$ and galaxy properties in the following sections.

Typically, either the \citet{Calzetti00} curve or the SMC curve is assumed for
the reddening of the stellar continuum in high-redshift galaxies. Because the SMC curve has a steeper intrinsic UV slope, these two dust attenuation curves lead to different $\xi_{ion}$, varying from less than 0.1 dex \citep[e.g.,][]{Bouwens16, Tang19} to more than 0.2 dex \citep[e.g.,][]{Shivaei18, Atek22}. Thus, we also perform SED fitting using the SMC curve for the stellar continuum, while still adopting the Milky Way curve \citep{Cardelli89} for fitting the nebular emission in this study.

\section{Results}
\label{sec:result}
\subsection{Star formation rates of low-mass $\mathrm{H\alpha}$ emitters}
\label{sec:pro}
\begin{figure*}
    \centering
    \vspace{-0.3cm}
    \includegraphics[width=0.49\textwidth]{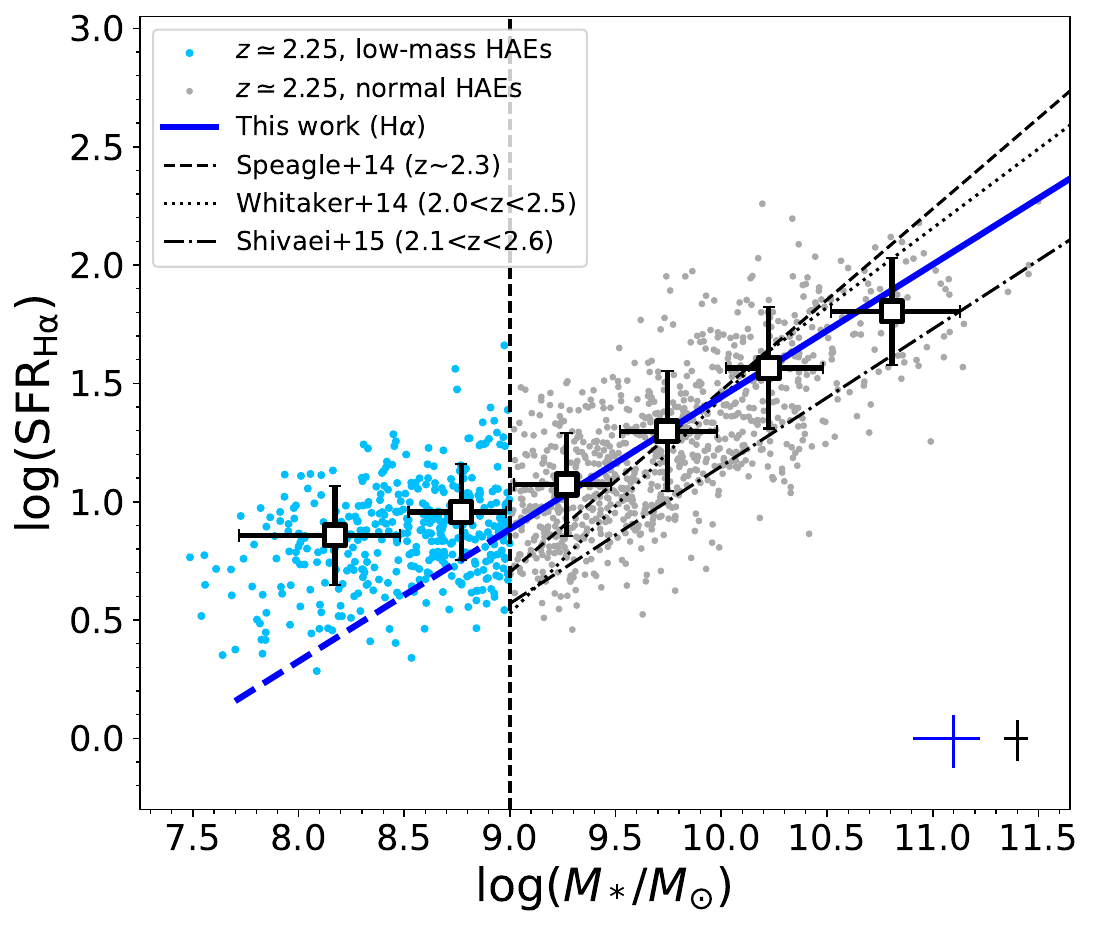}
    \includegraphics[width=0.49\textwidth]{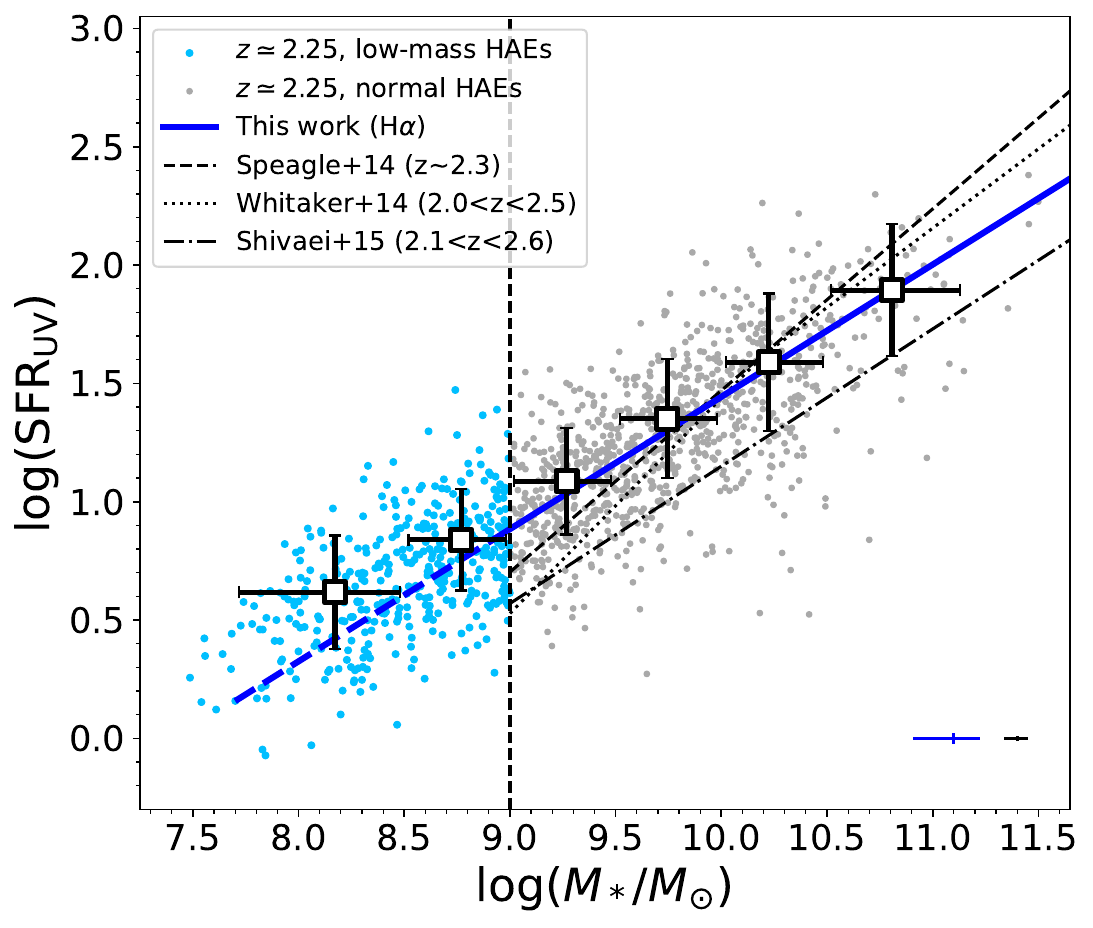}
    \label{fig:sfms}
    \caption{Based on the flux excesses in $K_s$ photometry, we define $\mathrm{H\alpha}$ emitters (HAEs) if they show $>3\sigma$ $\mathrm{H\alpha}$ detection in $K_s$ photometry. It is obvious that those low-mass HAEs ($<10^9M_{\odot}$) tend to scatter above the SFMS($\mathrm{H\alpha}$) but not obvious in SFMS(UV). Left: The star formation main sequence (SFMS) of 1318 HAEs at $z_{med} = 2.25$ in the ZFOURGE fields based on $\mathrm{H\alpha}$ emission line. Blue circles are 401 low-mass HAEs with $\mathrm{log(M_*/M_{\odot})} < 9$, while grey circles show the other HAEs in our catalog. Open squares are median stacks in 6 mass bins, while the error bars on them represent the scatter in each mass bin. Blue solid line is the best linear fit to the galaxies with $\mathrm{log}(M_*/M_{\odot})>9.0$, which is extrapolated to lower mass with blue dashed line. The best-fit SFMS from \citet{Whitaker2014}, \citet{Speagle2014} and \citet{Shivaei15} are also shown with black dotted, dashed, and dot-dashed lines, respectively. The error bars on the bottom-right corner represent the median uncertainty of low-mass HAEs (blue) and normal HAEs (black). Right: Same as the left panel, but SFR of each galaxy are calculated from the UV continuum. SFR($\mathrm{H\alpha}$) and SFR(UV) are corrected for dust attenuation using the Cardelli/Calzetti curve.}
\end{figure*}

In Figure \hyperref[fig:sfms]{1}, we present the the star formation main sequence (SFMS) of $\mathrm{H\alpha}$, $\mathrm{UV}$ for the HAEs from the C24 catalog. The low-mass HAEs are denoted as blue circles, while the median SFR in 6 mass bins is represented by large open circles. The mass bins are defined as follows: $\mathrm{log}(M_*/M_{\odot})<8.5$ for the first bin, $\mathrm{log}(M_*/M_{\odot})>10.5$ for the last bin, and the rest are divided into 0.5 dex widths. We apply the linear least squares regression to the SFR and stellar mass data points, above the mass completeness $\mathrm{log}(M_*/M_{\odot})=9.0$ \citep{Straatman16}. The best-fit linear correlation between SFRs and stellar masses, with the 68\% confidence interval on slope and intercept, is given by,

\begin{figure*}
    \centering
    \vspace{-0.3cm}
    \includegraphics[width=0.32\textwidth]{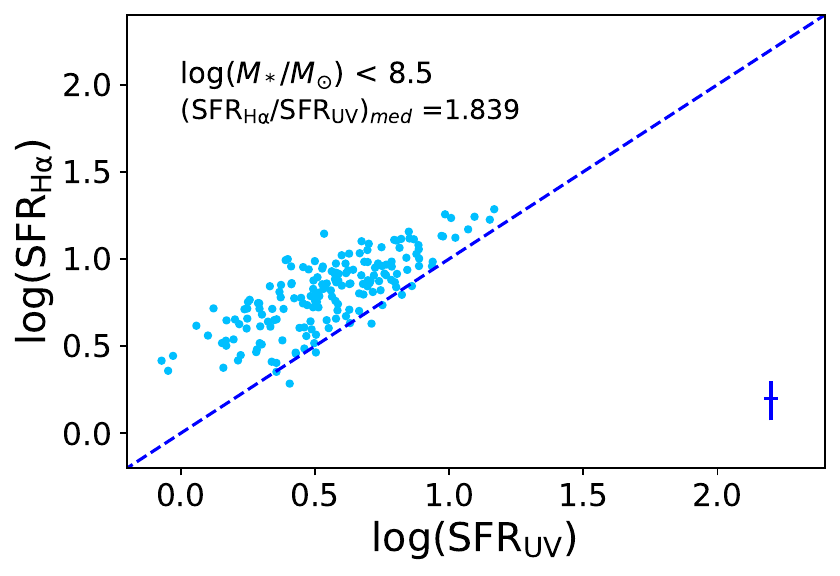}
    \includegraphics[width=0.32\textwidth]{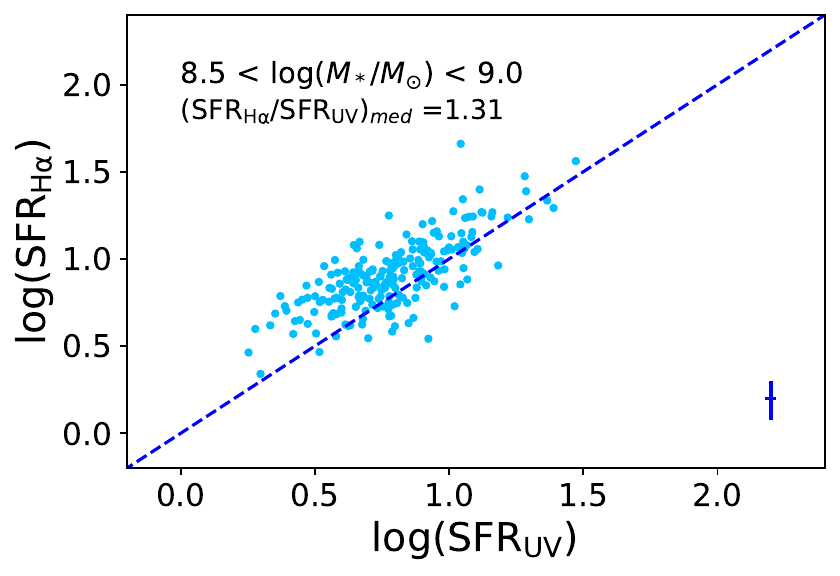}
    \includegraphics[width=0.32\textwidth]{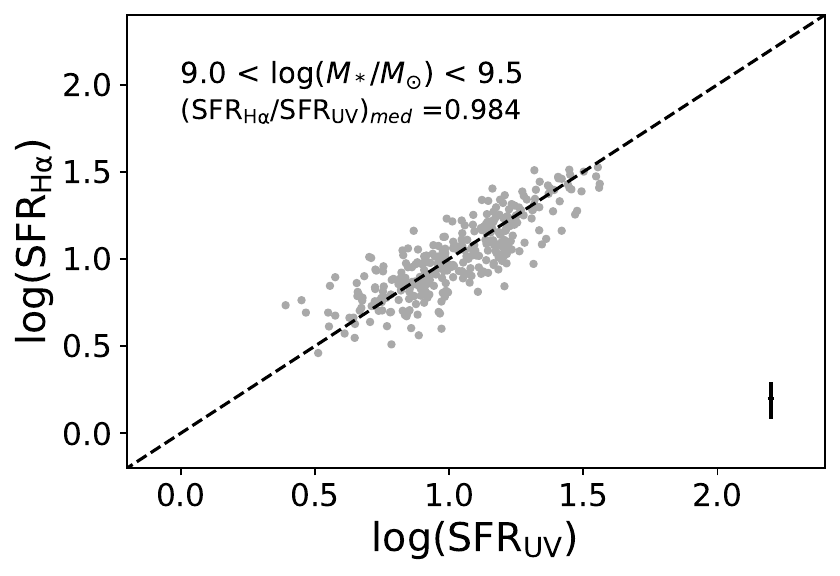}
    \includegraphics[width=0.32\textwidth]{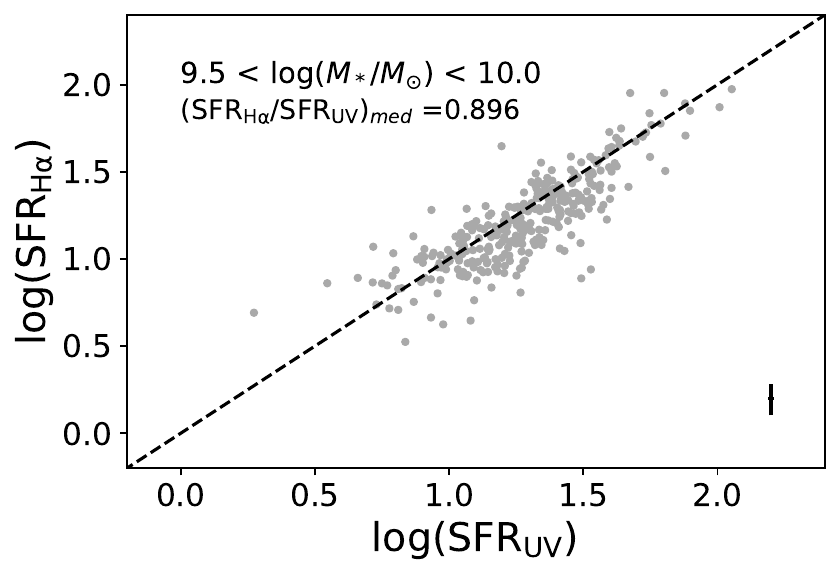}
    \includegraphics[width=0.32\textwidth]{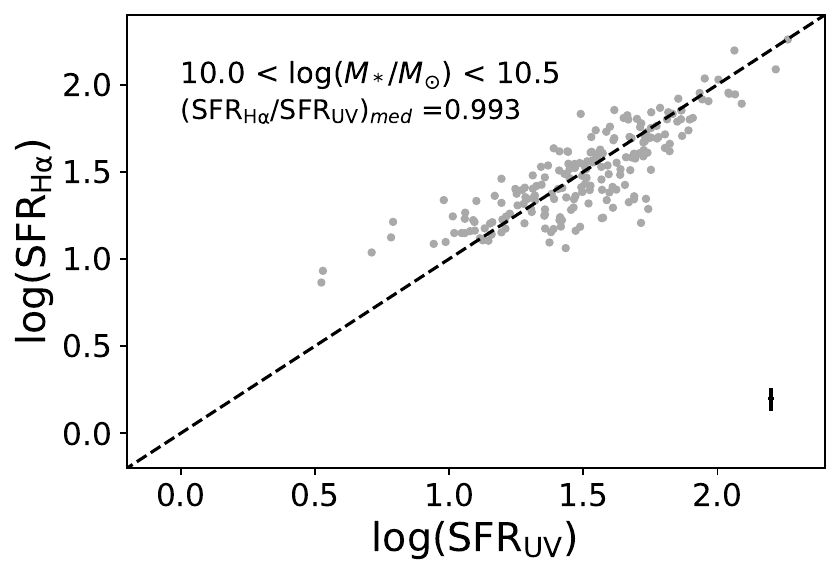}
    \includegraphics[width=0.32\textwidth]{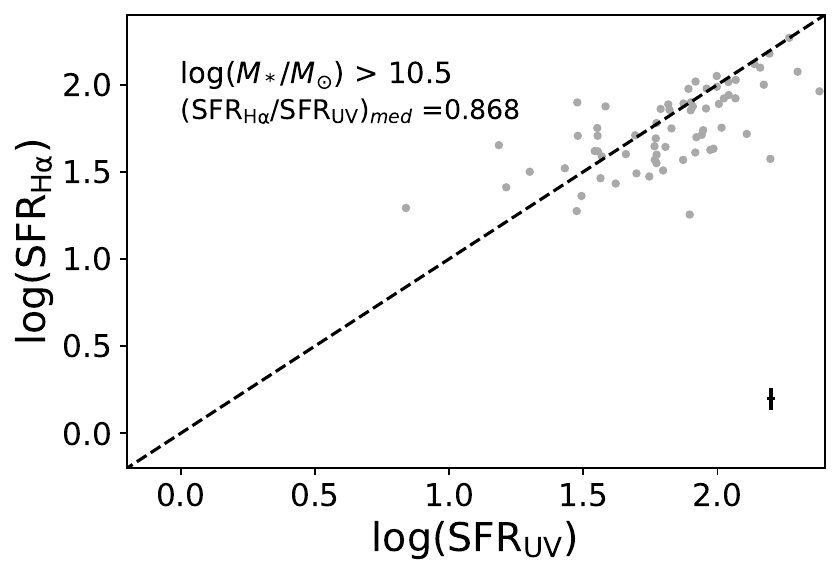}
    \label{fig:havsuv}
    \caption{Relation between the $\mathrm{SFR(H\alpha)}$ and $\mathrm{SFR(UV)}$ of the HAEs in our sample. SFR($\mathrm{H\alpha}$) and SFR(UV) are corrected for dust attenuation using the Cardelli/Calzetti curve. The color reﬂects different mass bins adopted in our study, as in Figure \hyperref[fig:sfms]{1}. More clearly, we find the low-mass HAEs ($<10^9M_{\odot}$) have the $\mathrm{SFR(H\alpha)}$/$\mathrm{SFR(UV)}$ ratio close to a factor of 2, indicating their characteristic galaxy properties.}
\end{figure*}
\begin{equation}
\label{equ:sfms}
\begin{aligned}
    \mathrm{log\,SFR}(\mathrm{H} \alpha) = & (0.56\pm0.03) \times \log M_* \\  & -(4.15\pm0.28);
    \\
    \mathrm{log\,SFR}(\mathrm{UV}) = & (0.60\pm0.04) \times \log M_* \\ &-(4.47\pm0.36).
\end{aligned}
\end{equation}

We extrapolate the $\mathrm{SFR}-M_*$ relation from Equation \hyperref[equ:sfms]{3} into low-mass domain ($<10^9M_{\odot}$) in Figure \hyperref[fig:sfms]{1}. In the right panel of Figure \hyperref[fig:sfms]{1}, the low-mass HAEs are found to be closer to the SFMS of UV, showing only a slight elevation in SFR(UV) compared to the SFMS by an average of 0.05 dex. However, a significant fraction of low-mass HAEs lie above the SFMS of $\mathrm{H\alpha}$, exhibiting an average SFR($\mathrm{H\alpha}$) higher than the SFMS by 0.25 dex. To further illustrate this trend, we plot the SFR($\mathrm{H\alpha}$) vs. SFR(UV) ratio for our sample in 6 mass bins in Figure \hyperref[fig:havsuv]{2}. In the high-mass domain ($>10^9M_{\odot}$), the ratio of these two SFR indicators is mainly close or below unity. In contrast, for galaxies with $\mathrm{log}(M_*/M_{\odot})<9.0$, the ratio is clearly above unity. We divide the HAEs in our study into two populations: 401 low-mass HAEs with $\mathrm{log}(M_*/M_{\odot})<9.0$ and 917 normal HAEs with $\mathrm{log}(M_*/M_{\odot})>9.0$.

The comparison of SFR($\mathrm{H\alpha}$) and SFR(UV) directly visualizes the burstiness of star formation activity. Previous studies \citep[e.g.,][]{Weisz12, Dominguez15, Emami19} have explored the time evolution of the $\mathrm{H\alpha}$-to-UV ratios in variety types of star formation history (SFH) models. Generally, SFR indicators have different timescales, with $\mathrm{SFR(UV)}$  having a timescale of $\sim$100 Myr and $\mathrm{SFR(H\alpha)}$ having a timescale of $\sim$10 Myr.
\citet{Emami19} has found that in galaxies undergoing rising star formation, the luminosity ratio between $\mathrm{H\alpha}$ and UV is expected to be higher than unity, and SFR($\mathrm{H\alpha}$) will reside above the SFMS by up to an order of magnitude. High-resolution hydrodynamical simulations by \citet{Sparre2017} have also shown that the scatter on the SFMS is larger for the $\mathrm{H\alpha}$-derived SFR because it is more sensitive to short bursts compared to the UV-based indicator.

\begin{figure}
    \includegraphics[width=1\linewidth]{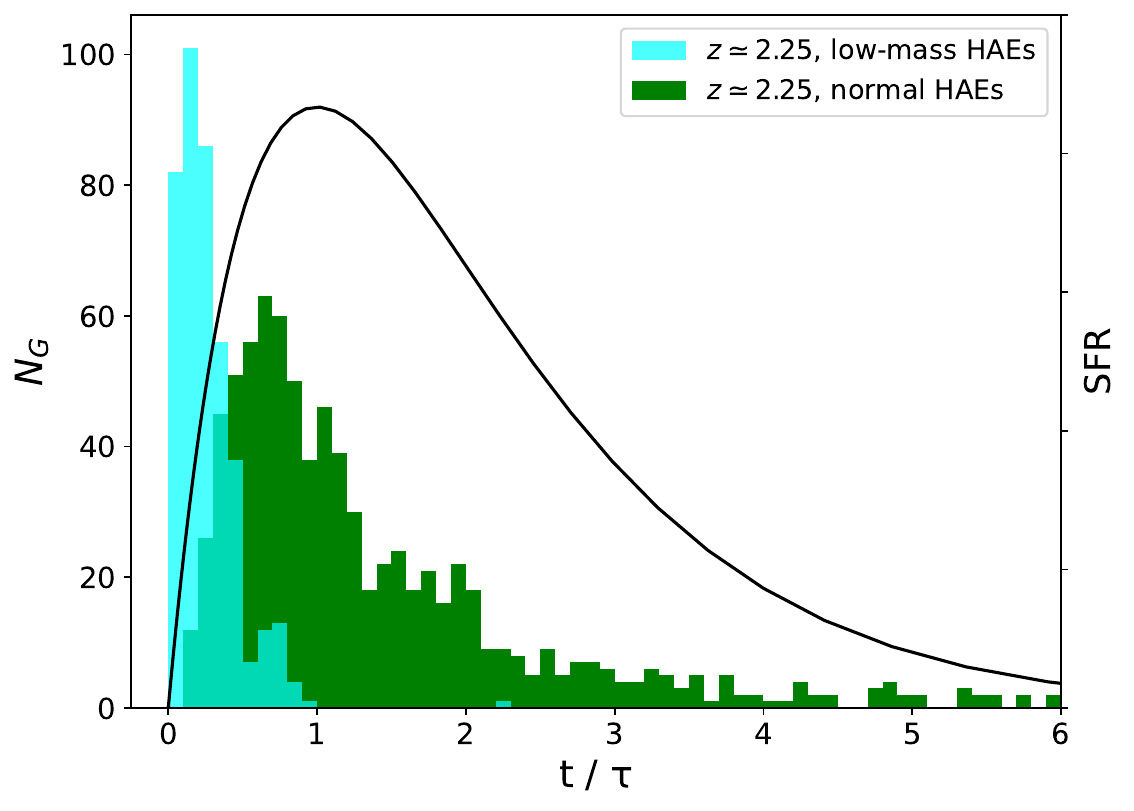}
    \label{fig:sfh}
    \vspace{-0.5cm}
    \caption{The star formation history (SFH) of the HAEs in our sample. SFH of the low-mass HAEs ($<10^9M_{\odot}$) and normal HAEs ($>10^9M_{\odot}$) are distributed as a histogram of t/$\tau$ in cyan column and green column. For reference, a model delayed-$\tau$ SFH is shown as black solid line. The SFH suggests that rising star-forming activities are occurring in these low-mass HAEs.}
\end{figure}

\begin{figure}
    \includegraphics[width=0.49\textwidth]{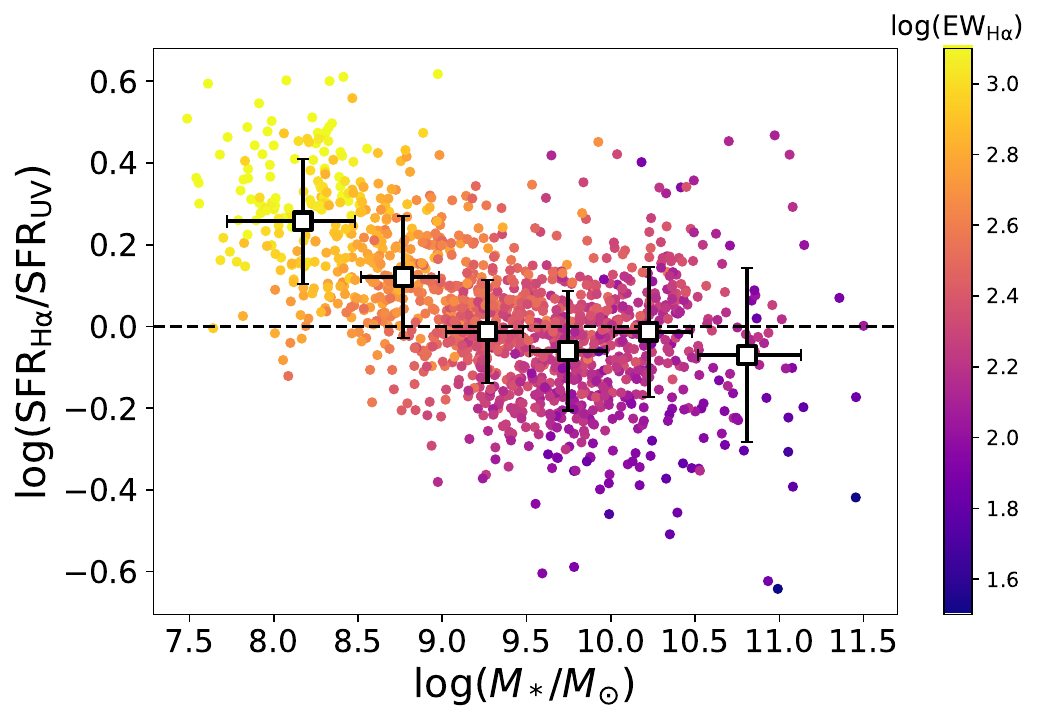}
    \label{fig:hauvew}
    \vspace{-0.5cm}
    \caption{The ratio of SFR($\mathrm{H\alpha}$) and SFR(UV) as a function of stellar mass. The color gradient of dots shows the $EW_{\mathrm{H\alpha}}$ of individual galaxy. Open squares are median stacks in 6 mass bins with the error bar representing the scatter in each mass bin.} Galaxies with high $EW_{\mathrm{H\alpha}}$ more than 1000$\mathrm{\AA}$ possess the highest SFR($\mathrm{H\alpha}$)/SFR(UV) and the lowest stellar masses in our sample.
\end{figure}

We further explore the SFH of the HAEs in our sample in Figure \hyperref[fig:sfh]{3}. In this study, we apply the delayed-$\tau$ models to model the SFH of galaxies as follows:
\begin{equation}
\label{equ:delaytau}
    \mathrm{SFR}(t) \propto \frac{t}{\tau^{2}} \times \exp \left(-t / \tau\right)\quad\mathrm{for}\ 0 \leq t \leq t_0.
\end{equation}
The distribution of the t/$\tau$ ratio of the HAEs reveals that the low-mass HAEs have a much smaller t/$\tau$, indicating that they are in their early stage of star-forming with rising SFRs. On the other hand, the distribution of normal HAEs shows more variability, indicating a range of SFHs in these galaxies. The rising star formation in low-mass galaxies indicates the abundance of young stellar population ($\leq 100\,\mathrm{Myr}$) in these systems, which can contribute to intense $\mathrm{H\alpha}$ emission lines. In Figure \hyperref[fig:hauvew]{4}, we plot the ratio of SFR($\mathrm{H\alpha}$) and SFR(UV) as a function of stellar mass, with an additional dimension of the equivalent width of $\mathrm{H\alpha}$ ($EW_{\mathrm{H\alpha}}$). It is not surprising that galaxies with high $EW_{\mathrm{H\alpha}}$ exceeding 1000 $\mathrm{\AA}$ exhibit the highest $\mathrm{SFR(H\alpha)}$/$\mathrm{SFR(UV)}$ratios. In comparison, galaxies with $EW_\mathrm{{H\alpha}}\sim100\mathrm{\AA}$ would follow the SFMS. 

As shown in Figure \hyperref[fig:hauvew]{4}, galaxy with the largest $EW_{\mathrm{H\alpha}}$ ($\mathrm{>1000\AA}$) have the lowest stellar masses in our sample with a mean value of $\mathrm{log}(M_*/M_{\odot})=8.09$. \citet{Stefanon22} conducted a stacking analysis of $z\sim8$ galaxies and found that the stacked galaxy exhibits extremely strong emission line, reaching up to $EW_{\mathrm{H\alpha}}\sim2000\mathrm{\AA}$. Besides, the stellar mass estimated from their stacked photometry is $\mathrm{log}(M_*/M_{\odot})=8.12$, which is very close to our result. Therefore, it is reasonable to assume that the low-mass HAEs in our study are possible analogs of galaxies during EoR.

\subsection{$\xi_{ion}$ evolution with galaxies properties}
\label{sec:xiion}
Here, we further investigate the ionizing photon production efﬁciency, $\xi_{ion}$, of the low-mass HAEs and their relationship to various galaxy properties. Since $f_{esc} = 0$ is assumed in our study, $\xi_{ion}$ is equivalent to the $\mathrm{SFR(H\alpha)}$/$\mathrm{SFR(UV)}$ ratio by amplifying $1.31\times10^{25}$. Thus,, $\xi_{ion}$ serves as a direct probe for the star formation activity in galaxies. 
$\xi_{ion}$ is also an important parameter in modeling the cosmic reionization and determining whether star-forming galaxies are capable of reionizing the universe \citep[e.g.,][]{Robertson13, Robertson15, Bouwens16}. In section \hyperref[sec:samp]{3}, we have applied two different recipes, Cardelli/Calzetti and Cardelli/SMC, for the nebular/continuum dust correction in our SED fitting analysis. We present the results from both approaches in the following sections, with the individual data points in the figures corresponding to the Cardelli/Calzetti curve.

\subsubsection{$\xi_{ion}$ and stellar mass}
\label{sec:xiionmass}
\begin{figure*}[hbt!]
    \centering
    \includegraphics[width=0.52\textwidth]{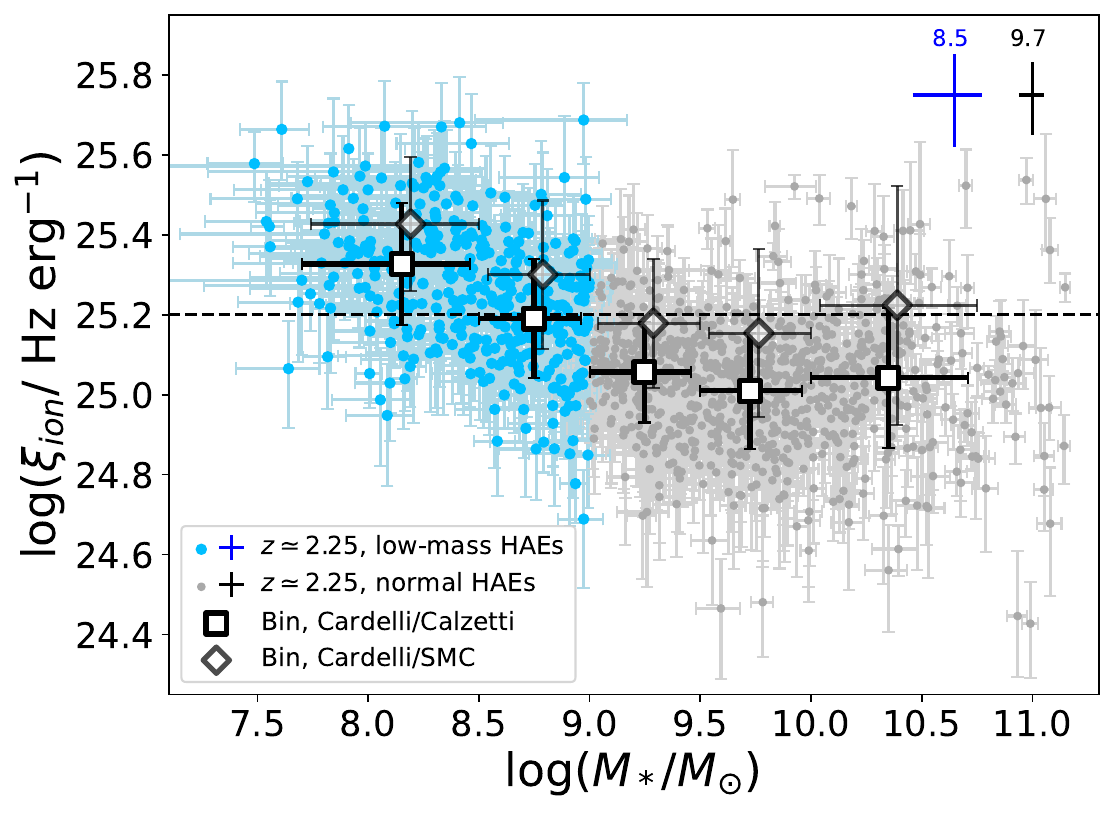}  
    \includegraphics[width=0.45\textwidth]{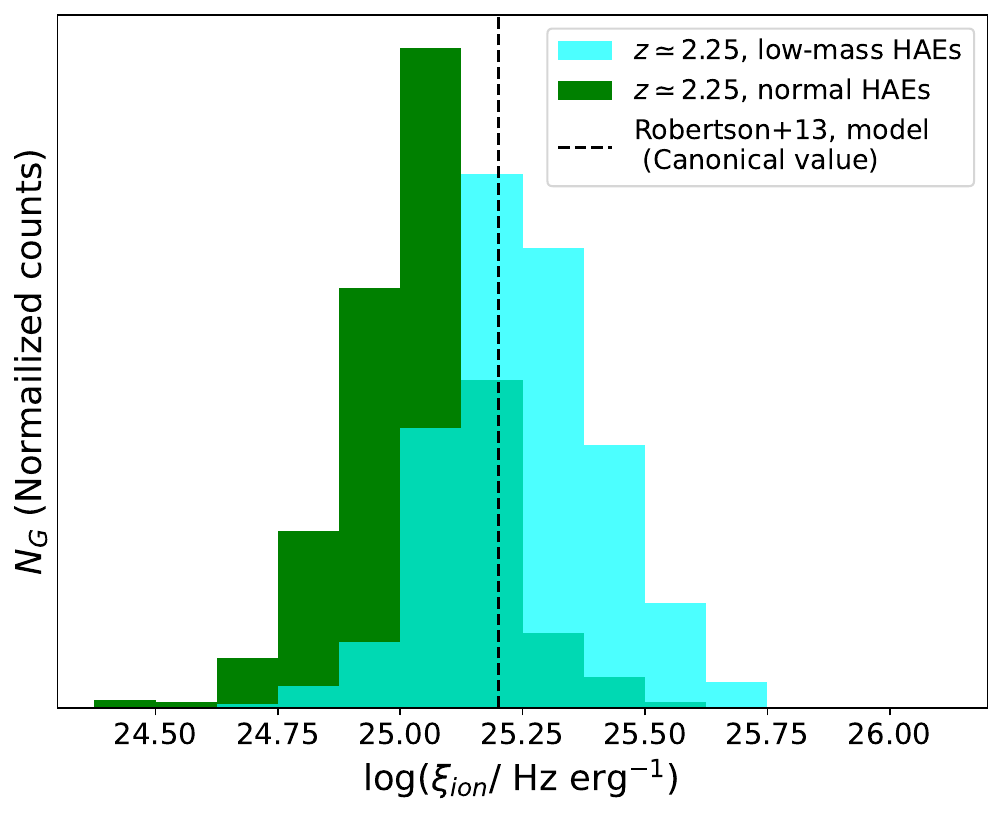}
    \label{fig:xiionmass}
    \caption{The ionizing photon production efficiency ($\xi_{ion}$) of all HAEs in our study. $\xi_{ion}$ is estimated from the $\mathrm{H\alpha}$ and UV luminosities followed by Equation \hyperref[equ:xiion]{1} and \hyperref[equ:xiion2]{2}. $\xi_{ion}$ of the low-mass HAEs is higher than other HAEs in our study by $\sim$0.2 dex, indicating a possible mass dependence. Left: Dependence of $\xi_{ion}$ on the stellar mass. The galaxy sample are separated into two populations, the normal HAEs and low-mass HAEs as in Figure \hyperref[fig:sfms]{1}. The error bars on the upper-right corner represent the median uncertainty of low-mass HAEs (blue) and normal HAEs (black), added with the median stellar mass of each subsample. Open squares and diamonds are median stacks in 5 mass bins, while the error bars on them represent the scatter in each mass bin. The square-shape stacks assume the Calzetti curve with for the UV dust correction, and the diamond-shape stacks assume an SMC curve. Right: The $\xi_{ion}$ distribution in the normal HAEs and low-mass HAEs assuming the Cardelli/Calzetti dust correction. The median values and errors for each population are $\mathrm{log}(\xi_{ion}) = 25.24^{+0.10}_{-0.13},\ 25.05^{+0.08}_{-0.10}$. The dashed lines in each figure indicate the canonical value of log($\xi_{ion}$)= 25.20 from \citet{Robertson13}.}
    \vspace{0.3cm}
\end{figure*}

We display the relationship between $\xi_{ion}$ and the stellar mass in our sample in the left panel of Figure \hyperref[fig:xiionmass]{5}. We also include open circles to represent the median $\xi_{ion}$ in different mass bins. To balance the number of samples in each bin, we combine the two most massive mass bins in Figure \hyperref[fig:sfms]{1}. The median $\xi_{ion}$ of the low-mass HAEs is $\mathrm{log}(\xi_{ion}/erg^{-1} Hz)=25.24^{+0.10}_{-0.13}\  (25.35^{+0.12}_{-0.15})$, assuming the Cardelli/Calzetti (Cardelli/SMC) curve. This result is very close to the O3Es subsample in \citet{Tang19} with $EW_{\mathrm{[O\pnt{III}]}}\simeq300-600\mathrm{\AA}$ (25.22).
On the other hand, the median $\xi_{ion}$ of the normal HAEs is $\mathrm{log}(\xi_{ion}/erg^{-1} Hz)=25.05^{+0.08}_{-0.10}\  (25.19^{+0.10}_{-0.13})$, which is quite similar to that of the galaxies from the
MOSDEF survey \citep[25.06;][]{Shivaei18}. When using an SMC extinction curve for the continuum, the $\xi_{ion}$ values are higher by $\sim$0.15 dex compared to the Calzetti curve.


The $\xi_{ion}$ distributions of low-mass HAEs and normal HAEs are exhibited in the right panel of Figure \hyperref[fig:xiionmass]{5}. The intrinsic scatter, represented by the standard deviation of the distribution, is $0.16\,(0.19)$ dex for low-mass HAEs assuming the Cardelli/Calzetti (Cardelli/SMC) curve. For normal HAEs, the intrinsic scatter is $0.16\,(0.26)$ dex.

We find that, on average, the low-mass HAEs have $\sim$0.2 dex higher $\xi_{ion}$ compared to the normal HAEs, indicating a higher efficiency of producing ionizing photons in the low-mass galaxies. 
In the high-mass domain, we observe no significant evolution of $\xi_{ion}$ with stellar mass. This trend is consistent with \citet{Shivaei18} at $z\sim2$ using MOSDEF spectroscopic data. The $\xi_{ion}$ values from MOSDEF remain relatively constant down to $10^{9.5}M_{\odot}$ but show an increase in the lowest mass bin of $10^9M_{\odot}$, suggesting a possible mass dependence of $\xi_{ion}$. Our method successfully extends the mass range and provides evidence for the existence of a mass dependence of $\xi_{ion}$ below $10^9M_{\odot}$. In contrast, at similar redshift, \citet{Emami20} has found that $\xi_{ion}$ is generally independent of the stellar mass in their sample of 28 lensed dwarf galaxies, which span the range of $8.0<\mathrm{log}(M_*/M_{\odot})<10.0$. Their sample exhibits a nearly twice as large intrinsic scatter compared to our study. It is worth noting that high lensing magnification can also introduce significant magnification differences across the galaxy sample, leading to uncertainties when measuring $\xi_{ion}$. Differences in sample selection and observational techniques may contribute to the discrepancy found in these studies.

\subsubsection{$\xi_{ion}$ and UV Properties}
\begin{figure*}[hbt!]
    \centering
    \includegraphics[width=0.49\textwidth]{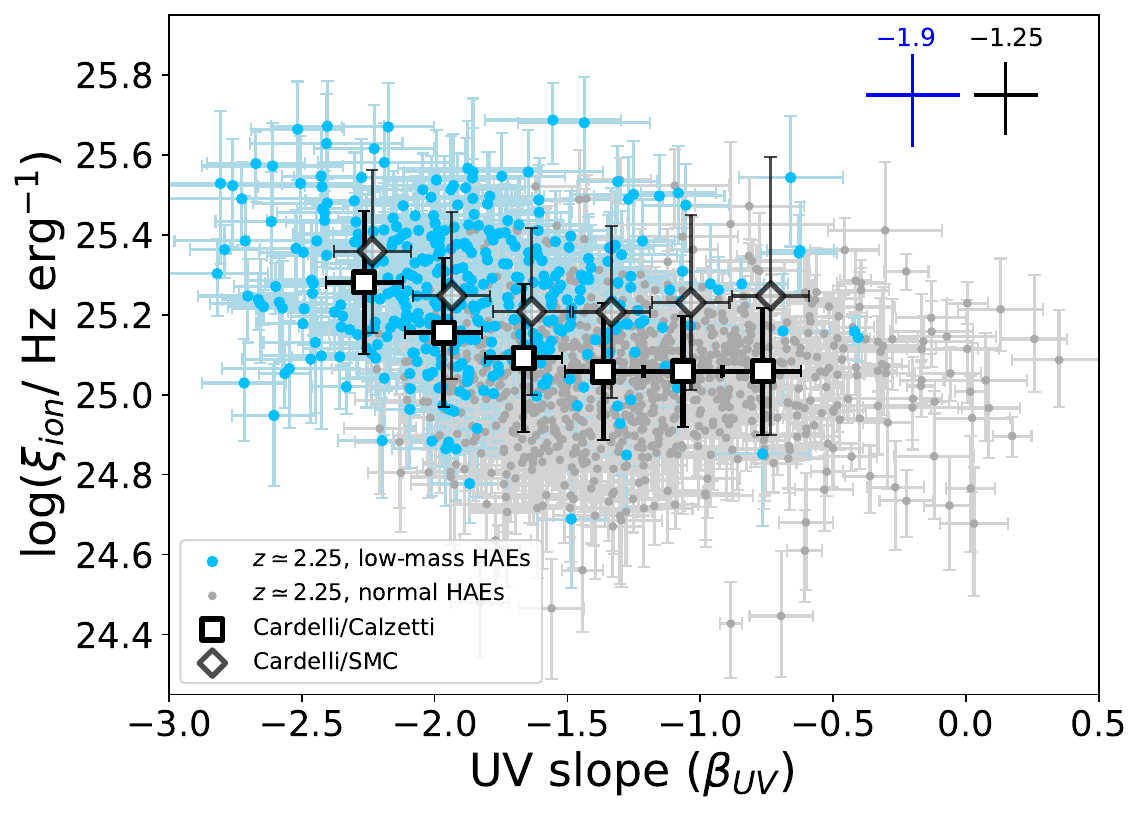}
    \includegraphics[width=0.49\textwidth]{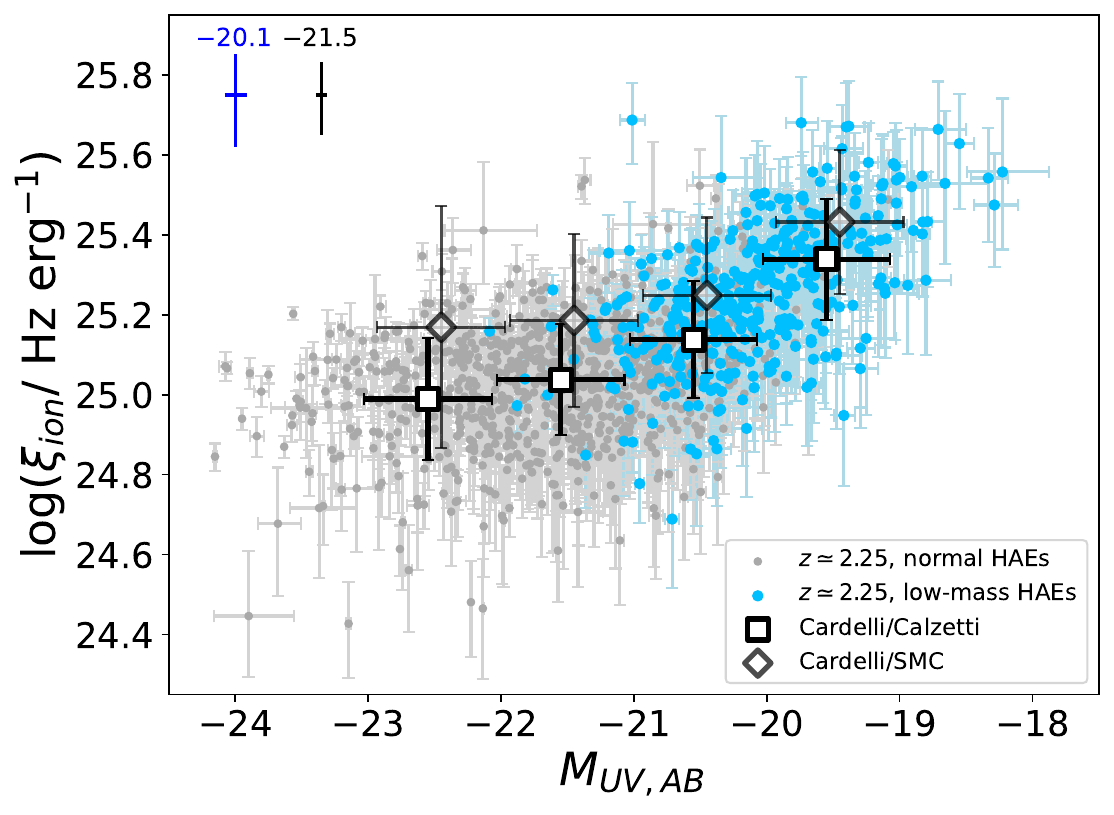}
    \label{fig:xiionvspro}
    \caption{$\xi_{ion}$ as a function of the UV spectral slope ($\beta_{\mathrm{UV}}$, Left), the absolute UV magnitude at 1500$\mathrm{\AA}$ ($M_{\mathrm{UV}}$, Right). Symbols are as in Figure \hyperref[fig:xiionmass]{5}. Open squares and diamonds are median stacks for these two galaxy properties in Table \hyperref[tab:xiionuv]{2}, but represent different curve for the UV dust correction, while the error bars on them are the scatter in each bin. 
    The error bars on the corner represent the median uncertainty of low-mass HAEs (blue) and normal HAEs (black), added with the number value of median UV slope and UV magnitude.
    It is clear that galaxies with bluer $\beta_{\mathrm{UV}}$ and fainter UV luminosity are likely to hold larger $\xi_{ion}$ than others, encouraging that these galaxies would play an important role in reionizing the universe.}
    \vspace{0.3cm}
\end{figure*}

It has been suggested that UV-faint galaxies could be significant contributors to cosmic reionization, based on several observations. One reason is the higher number density, indicated by the steep slope at the faint-end of the UV luminosity function \citep[e.g.,][]{Atek15, Bouwens15}. Another factor is the potentially higher LyC escape fraction for faint galaxies \citep{Grazian17}. Investigating the UV properties of the low-mass HAEs can provide valuable insights into EoR. We report our results in Figure \hyperref[fig:xiionvspro]{6} and Table \hyperref[tab:xiionuv]{2}.

\begin{table}[hbt!]
    \small
    \centering
    \label{tab:xiionuv}
    \caption{Median $\xi_{ion}$ for HAEs at $z\sim2.3$ separated into different bins of stellar masses, UV luminosities and UV-slopes.}
    \begin{tabular}{cccc}
    \hline \hline & & \multicolumn{2}{c}{$\log \bar{\xi}_{ion} /\left[\mathrm{Hz}\,\mathrm{erg}{ }^{-1}\right]$} \\
    \cline { 2 - 4 } (Sub)sample & \multicolumn{1}{c}{$N_{gal}$} & Calzetti$\mathrm{\,^{a}}$ & SMC \\
    \hline
    All HAEs$\mathrm{\,^{b}}$ & 1318 & $25.09_{-0.11}^{+0.09}$ & $25.24_{-0.14}^{+0.11}$ \\
    \hline
    low-mass HAEs$\mathrm{\,^{c}}$ & 401 & $25.24_{-0.13}^{+0.10}$ & $25.35_{-0.15}^{+0.12}$\\
    Normal HAEs & 917 & $25.05_{-0.10}^{+0.08}$ & $25.19_{-0.13}^{+0.10}$\\
    \hline
    $\mathrm{log}(M_*/M_{\odot})>10.0$ & 270 & $25.04_{-0.08}^{+0.07}$ & $25.21_{-0.12}^{+0.10}$ \\
    $9.5<\mathrm{log}(M_*/M_{\odot})<10.0$ & 326 & $25.01_{-0.10}^{+0.08}$ & $25.16_{-0.13}^{+0.10}$ \\
    $9.0<\mathrm{log}(M_*/M_{\odot})<9.5$ & 321 & $25.06_{-0.12}^{+0.10}$ & $25.18_{-0.14}^{+0.11}$ \\
    $8.5<\mathrm{log}(M_*/M_{\odot})<9.0$ & 221 & $25.19_{-0.13}^{+0.10}$ & $25.30_{-0.16}^{+0.12}$\\
    $\mathrm{log}(M_*/M_{\odot})<8.5$ & 180 & $25.33_{-0.13}^{+0.11}$ & $25.43_{-0.15}^{+0.12}$\\
    \hline
    $\beta_{\mathrm{UV}}<-2.1$ & 121 & $25.28_{-0.13}^{+0.10}$ & $25.36_{-0.15}^{+0.11}$ \\
    $-2.1<\beta_{\mathrm{UV}}<-1.8$ & 212 & $25.16_{-0.13}^{+0.10}$ & $25.25_{-0.15}^{+0.11}$ \\
    $-1.8<\beta_{\mathrm{UV}}<-1.5$ & 265 & $25.09_{-0.12}^{+0.09}$ & $25.21_{-0.14}^{+0.11}$ \\
    $-1.5<\beta_{\mathrm{UV}}<-1.2$ & 306 & $25.06_{-0.11}^{+0.09}$ & $25.21_{-0.14}^{+0.11}$ \\
    $-1.2<\beta_{\mathrm{UV}}<-0.9$ & 194 & $25.06_{-0.09}^{+0.08}$ & $25.23_{-0.13}^{+0.10}$ \\
    $\beta_{\mathrm{UV}}>-0.9$ & 220 & $25.06_{-0.07}^{+0.07}$ & $25.24_{-0.14}^{+0.11}$ \\
    \hline
    $M_{\mathrm{UV}}<-22.0$ & 284 & $24.99_{-0.06}^{+0.06}$ & $25.16_{-0.09}^{+0.08}$ \\
    $-22.0<M_{\mathrm{UV}}<-21.0$ & 441 & $25.04_{-0.10}^{+0.08}$ & $25.19_{-0.14}^{+0.11}$ \\
    $-21.0<M_{\mathrm{UV}}<-20.0$ & 405 & $25.14_{-0.13}^{+0.10}$ & $25.25_{-0.15}^{+0.11}$ \\
    $M_{\mathrm{UV}}>-20.0$ & 188 & $25.34_{-0.15}^{+0.11}$ & $25.43_{-0.16}^{+0.12}$ \\
    \hline
    \end{tabular}
    \begin{tablenotes}
    \item \textbf{Notes.} Measurments of the UV properties are only based on the observed photometry mentioned in section \hyperref[sec:messfr]{3.3}, which have no relationship to the SED fittings. \\${ }^{\text {a }}$The uncertainty on each number is the median uncertainty of the subsample, which is different from the scatter shown in Figure \hyperref[fig:xiionvspro]{10}. \\
    ${ }^{\text {b }}$AGNs and QGs (the total number is 40) are excluded when deriving $\xi_{ion}$.\\
    ${ }^{\text {c }}$The low-mass HAEs are those with stellar mass $<10^9M_{\odot}$, while the normal HAEs are others.
    \end{tablenotes}
\end{table}

The left panel of Figure \hyperref[fig:xiionvspro]{6} shows the relationship between $\xi_{ion}$ and UV slope ($\beta_{\mathrm{UV}}$) for our sample. We observe a gradual increase in $\xi_{ion}$ for galaxies with $\beta_{\mathrm{UV}}<-2.0$, with the bluest $\beta_{\mathrm{UV}}$ objects showing an elevation of more than 0.2 dex. The UV slope is sensitive to the age of stellar population of massive stars (O- and B- type stars) in a galaxy. In ionization-bounded photoionization models \citep[e.g.,][]{Topping22}, the bluest expected UV slope is $\beta_{\mathrm{UV}}\simeq-2.6$ in a dust-free case with stellar age of $<30\,\mathrm{Myr}$. An older stellar population would lead to a redder UV slope up to $\beta_{\mathrm{UV}}\sim-2.0$ and less production of ionizing photons, resulting in a lower $\xi_{ion}$. Otherwise, $\xi_{ion}$ remains nearly unchanged for galaxies with redder $\beta_{\mathrm{UV}}$ values ($>-1.5$). The effect of dust attenuation might become dominant in this region. Applying the SMC correction for the UV continuum leads to a similar result, albeit with a slightly lower elevation of $\sim$0.15 dex.

Our results are consistent with the literature results from \citet{Shivaei18} at similar redshift and \citet{Bouwens16} at $z\simeq4-5$, both of which inferred an elevated $\xi_{ion}$ in galaxies with the bluest $\beta_{\mathrm{UV}}$. On the other hand, studies by \citet{Emami20} and \citet{Onodera20} have found no correlation between $\xi_{ion}$ and $\beta_{\mathrm{UV}}$.
As mentioned above, \citet{Emami20} used a sample of 28 lensed dwarf galaxies, with rest-frame equivalent widths up to $1500\mathrm{\AA}$. Also, the results from \citet{Onodera20} are based on $\sim20$ extreme O3Es with rest-frame equivalent widths up to $2000\mathrm{\AA}$. We speculate that the discrepancy between our results and previous studies may be attributed to the selection biases, given that their samples encompass a small number of extreme emitters, whereas our sample is more inclusive.

The low-mass HAEs in our study have a median value of $\beta_{\mathrm{UV}} = -1.90$, much higher than normal HAEs by $\sim$0.7. Comparing with the dust-free value of $\beta_0=-2.23$ from the \citet{Meurer99} calibration, we still have 72 low-mass HAEs (more than 1/6) with the bluest $\beta_{\mathrm{UV}} < -2.23$. Such blue $\beta_{\mathrm{UV}}$ values suggest a very young stellar population with minimal dust content in the system.

Next, we explore the relationship between $\xi_{ion}$ and UV absolute magnitude ($M_{\mathrm{UV}}$) in the right panels of Figure \hyperref[fig:xiionvspro]{6}. 
In our sample, we observe an increase in $\xi_{ion}$ for the faintest galaxies compared to the brighter ones, with a difference of more than 0.3 dex. This trend suggests a dependence between $\xi_{ion}$ and $M_{\mathrm{UV}}$ for the HAEs in our study. These results are similar to those observed in Lyman-alpha emitters (LAEs) at $z\sim3$ from \citet{Nakajima18, Nakajima20} with the faint end of UV magnitude to $M_{\mathrm{UV}}\simeq-19.5$ mag. On the other hand, \citet{Shivaei18} and \citet{Emami20} did not find a strong correlation between these two parameters in their studies. While, it should be noted that the galaxy sample in \citet{Shivaei18} has been selected through spectroscopy, which may introduce a bias towards brighter objects $M_{\mathrm{UV}}\leq-21$ mag. It can be also inferred from our results that brighter galaxies exhibit a weaker dependence between $\xi_{ion}$ and $M_{\mathrm{UV}}$.

\begin{figure*}
    \centering
    \includegraphics[width=0.49\textwidth]{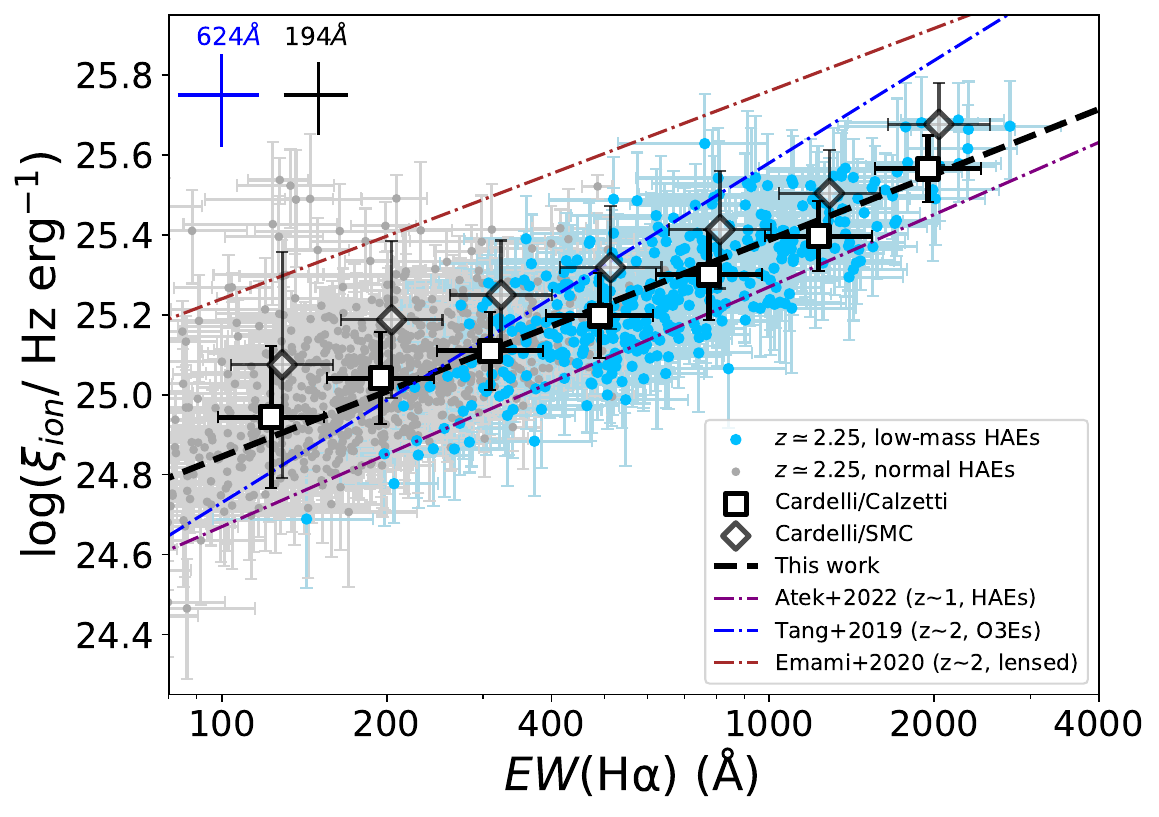}
    \includegraphics[width=0.49\textwidth]{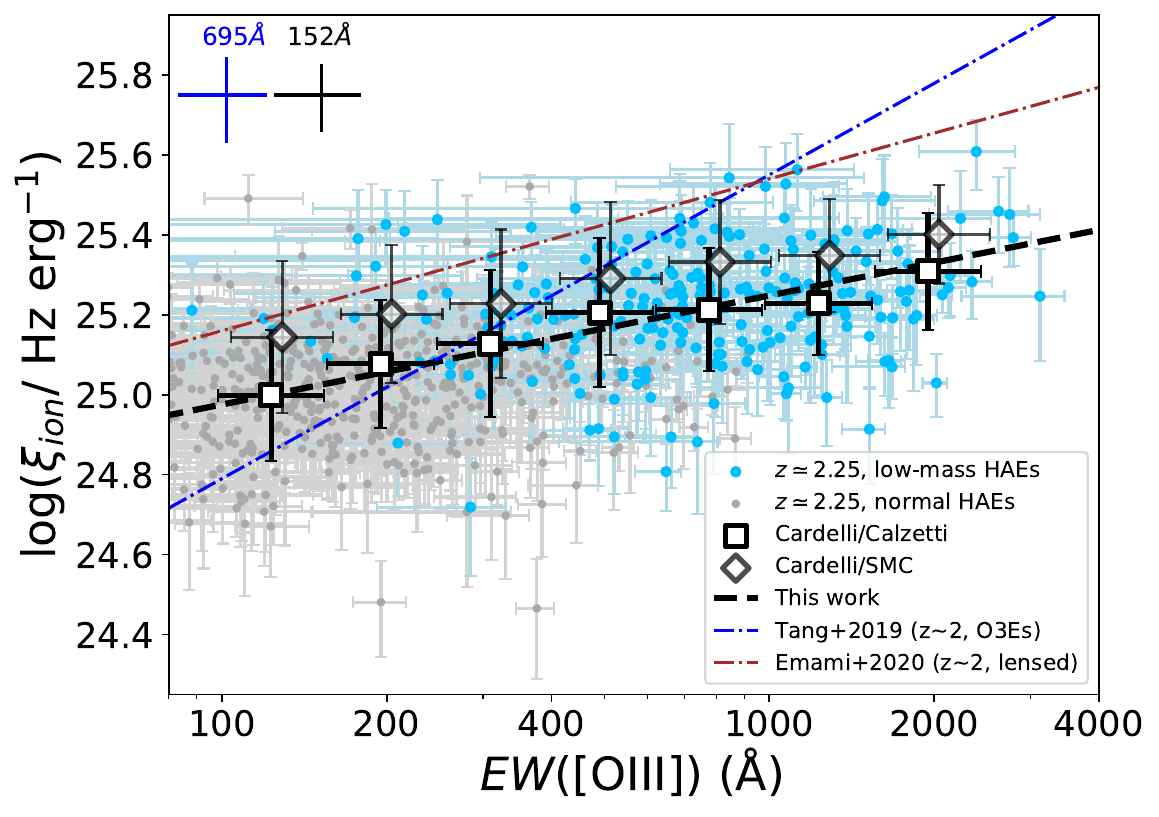}
    \label{fig:xiionvsline}
    \vspace{-0.1cm}
    \caption{$\xi_{ion}$ as a function of the equivalent width of $\mathrm{H\alpha}$ ($EW_{\mathrm{H\alpha}}$, Left), equivalent width of $\mathrm{[O\pnt{III}]}$ ($EW_{\mathrm{[O\pnt{III}]}}$, Right). $\mathrm{[O\pnt{III}]}$ are derived from the flux excesses in $H_s$/$H_l$ photometry. Symbols are as in Figure \hyperref[fig:xiionmass]{5}. Open squares and diamonds are median stacks for various galaxy properties, but represent different curve for the UV dust correction. The error bars on the corner represent the median uncertainty of low-mass HAEs (blue) and normal HAEs (black), added with the number value of median $EW_{\mathrm{H\alpha}}$ and $EW_{\mathrm{[O\pnt{III}]}}$. For both panels, the best-fit relation between $\xi_{ion}$ and $EW_{\mathrm{H\alpha}}$ ($EW_{\mathrm{[O\pnt{III}]}}$) is shown with a black dashed line where we use the Calzetti curve for dust correction. Other relationship derived for HAEs at $0.7 < z < 1.5$ \citep{Atek22}, $\mathrm{[O\pnt{III}]}$ emitters at $1.3 < z < 2.4$ \citep{Tang19}, lensed low-mass galaxies at $z\sim2$ \citep{Emami20} are shown with purple, blue, brown dashed-dotted line, respectively. Note that \citet{Emami20} and \citet{Atek22} used the SMC curve for dust correction, which results in a higher estimation of $\xi_{ion}$.}
    \vspace{0.3cm}
\end{figure*}

\subsubsection{$\xi_{ion}$ and nebular emission lines}
The optical nebular emission lines in galaxies provide a wealth of information  on the physical parameters, including the stellar population, chemical abundance, and ionization parameter, which may also be related to $\xi_{ion}$.

As $\xi_{ion}$ serves as a useful indicator of the stellar populations in galaxies, with younger populations contributing more to the $\mathrm{H\alpha}$ emission lines, we expect to find a universal relationship between $\xi_{ion}$ and the equivalent width of $\mathrm{H\alpha}$. This relationship is depicted in the left panel of Figure \hyperref[fig:xiionvsline]{7}. Assuming the Cardelli/Calzetti curve, we fit a linear relationship between these two attributes:
\begin{equation}
    \label{equ:haxiion}
    \begin{aligned}
    \log \xi_{ion}= & (0.54\pm0.03) \times \log(EW_{\mathrm{H\alpha}}) \\ & + (23.76\pm0.09).
    \end{aligned}
\end{equation}
For the Cardelli/SMC curve, we find a relationship with a slope of $0.51\pm0.04$ and an intercept of $23.96\pm0.11$.

Since large [O{\sc iii}] equivalent widths are typically produced by massive stellar populations with sub-solar metallicities, which meanwhile produce large amounts of ionizing photons, it is suggested that there also exists a correlation between $\xi_{ion}$ and $EW_{\mathrm{[O\pnt{III}]}}$. This correlation has been observed in both local star-forming galaxies \citep{Chevallard18} and high-redshift emitters \citep{Tang19, Emami20, Nakajima20}, indicating that systems with higher $EW_{\mathrm{[O\pnt{III}]}}$ are more efficient in producing ionizing photons. Our results also indicate a similar relationship between these two attributes, assuming the Cardelli/Calzetti curve:
\begin{equation}
    \label{equ:o3xiion}
    \begin{aligned}
    \log \xi_{ion}= & (0.27\pm0.04) \times \log(EW_{\mathrm{[O\pnt{III}]}}) \\ & + (24.43\pm0.11).
    \end{aligned}
\end{equation}
Similarly, for the Cardelli/SMC curve, we find a relationship with a slope of $0.24\pm0.04$ and an intercept of $24.62\pm0.10$.

For reference, we also overlay the best-fitting relations from \citet{Tang19}, \citet{Emami20} and \citet{Atek22} in Figure \hyperref[fig:xiionvsline]{7}. Although our sample shows a similar trend to literature, we find a clear discrepancy in terms of the slopes and intercepts. One possible reason for this discrepancy could be the treatment of dust extinction. \citet{Tang19} has derived the relationship for their [O{\sc iii}] emitters (O3Es) using both the Calzetti law and the SMC law. Similar to our sample, applying the SMC law results in a similar slope but an elevation in the intercept by $\sim0.15$ dex. The difference in galaxy samples may also contribute to the discrepancy in slopes. \citet{Emami20} has suggested that the slope in $\log \xi_{ion}$ and $\log EW_{\mathrm{H\alpha}}$ ($\log EW_{\mathrm{[O\pnt{III}]}}$) becomes shallower at lower equivalent widths. To explore this further, we derived the best-fitting result only for HAEs with $EW_{\mathrm{H\alpha}}\ (EW_{\mathrm{[O\pnt{III}]}}) > 250\mathrm{\AA}$. Both curves shows a steeper slope of $0.61\pm0.03$ ($0.32\pm0.05$), supporting the idea in \citet{Emami20}. While, even when considering larger equivalent widths and fitting the galaxies accordingly, the discrepancy with \citet{Tang19} still remains. The difference in sample selection, with \citet{Tang19} focusing on the most intense O3Es, may contribute to the discrepancy in the best-fitting relationship.

In any case, it is worth noting that the relationship between $\xi_{ion}$ and the equivalent widths of nebular emission lines, such as H$\alpha$ and [O{\sc iii}], is observed globally at $z\sim2$. These strong nebular emission lines can serve as proxies for measuring $\xi_{ion}$.

\subsection{$\xi_{ion}$ evolution with redshift}
\label{sec:xiionz}
\begin{figure*}[hbt!]
    \centering
    \includegraphics[width=0.9\textwidth]{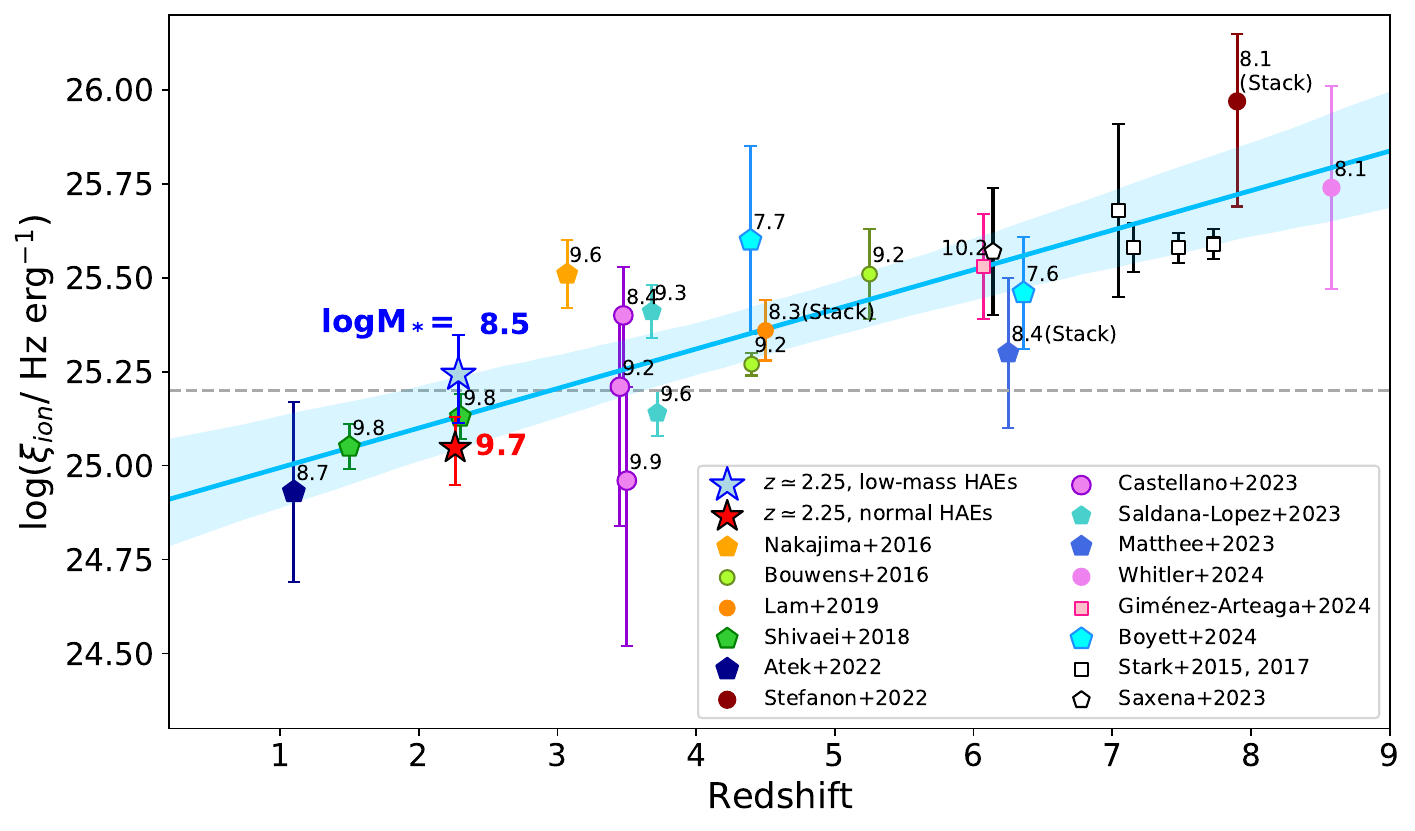}
    \label{fig:xiionz}
    \vspace{-0.2cm}
    \caption{Evolution of estimated ionizing photon production efficiency ($\xi_{ion}$) with redshift. Our measurement at $z \sim 2.25$ (blue and red stars) are compared with literature results at a wide range of redshifts, include \citet{Stark15, Stark17, Bouwens16, Nakajima16, Shivaei18, Atek22, Stefanon22, Castellano23,Saldana-Lopez23,Matthee22, Saxena23, Whitler23, Arteaga24, Boyett24}. The spectroscopic-selected samples are marked as pentagonal symbols, the photometric-selected sample as circular ones and the individual sample as square ones. If multiple values exist in the literature, we take the result based on the Calzetti curve for UV continuum. The numbers next to the markers are the average stellar mass from each sample, and those without stellar mass are marked as open symbols. The HAEs from our sample are marked with different colors. The dashed horizontal line indicates the canonical value of log($\xi_{ion}$)= 25.20 \citep{Robertson13}. The blue solid line and the shaded area show the best-fitting linear regression results to all data with 95\% confidence interval. We claim a evolution of $\xi_{ion}$ with lookback time, with the largest sample at $z\sim2$ so far. This evolution can be explained by more bursty star formation on average at higher redshifts.}
    \vspace{0.2cm}
\end{figure*}

Constraining $\xi_{ion}$ during EoR is an important task for cosmic reionization models. Measuring $\xi_{ion}$ for a large number of low-mass galaxies at $z > 6$ remains very challenging. \citet{Shivaei18} has suggested a possible evolution of $\xi_{ion}$ with redshift, which could provide insights for extrapolating $\xi_{ion}$ to higher redshift. 
We compare our results with previous studies of $\xi_{ion}$ at various redshifts in Figure \hyperref[fig:xiionz]{8}. Note that the galaxy samples included here are compiled in a heterogeneous manner, encompassing continuum-selected, stacked, and emission line-selected objects, which might introduce additional scatter.

\citet{Atek22} investigated $\xi_{ion}$ of a large sample of 1167 HAEs at $0.7 < z < 1.5$ by using 3D-HST grism and imaging data. More than half of their 3D-HST sample are low-mass galaxies below $10^9M_{\odot}$, which have a similar stellar mass range to our sample. Since the low-mass 3D-HST galaxies are also the highest-EW galaxies, we consider them as analog sample at lower redshift.

\citet{Shivaei18} used the spectroscopic and UV imaging data from the MOSDEF survey to measure $\xi_{ion}$ at $z\sim1.5$ and $z\sim2.3$ of 673 galaxies. The lowest stellar mass bin of \citet{Shivaei18} is $10^{9.3}M_{\odot}$, lacking galaxies with $\mathrm{log}(M_*/M_{\odot})<9.0$. The large number of low-mass HAEs found in our study are essential to complete the whole picture at $z\sim2$.

\citet{Nakajima16} used the spectroscopic measurement of H$\beta$ emission line and
SED-derived UV continuum to infer $\xi_{ion}$ of 13 LAEs as well as 2 Lyman Break Galaxies (LBGs) at $z\simeq3.1-3.7$. Their LAE sample consists of less dusty system with color excess $E(B-V)$ almost consistent with zero. Also, the large [O{\sc iii}]/[O{\sc ii}] ratio indicates the higher ionization properties of their LAE sample. 

\citet{Bouwens16} measured the H$\alpha$ emission line from the observed IRAC fluxes in $\sim300$ galaxies at $z\simeq3.8-5.0$ and 22 galaxies at $z\simeq5.1-5.7$. They calculated H$\alpha$ flux through the flux excesses in IRAC data, i.e, the [3.6]-[4.5] color. Besides, their result showed that assuming the SMC extinction law led to $\sim0.1$ dex higher $\xi_{ion}$ than that with the Calzetti one.

\citet{Lam19} performed a similar estimation of $\xi_{ion}$ from the flux excesses in IRAC as \citet{Bouwens16} but by using the stacked IRAC colors in a sample of galaxies. Comparing to \citet{Bouwens16}, \citet{Lam19} extended $\xi_{ion}$ measurements to even lower UV luminosities. The distribution of stellar mass and the large number sample in \citet{Lam19} was comparable to the low-mass HAEs in our study, indicating the similarity of the galaxy samples.

\citet{Stark15,Stark17} discussed 4 individual, bright, lensed galaxies at $z\sim7$. $\xi_{ion}$ is inferred from stellar population
and photo-ionization models. Significant detection of FUV emission lines of these galaxies, including Ly$\alpha$, 
Similar to the sample from \citet{Nakajima16}

\citet{Stefanon22} measured the median-stacked galaxy properties of 102 LBGs at $z\sim8$. H$\alpha$ flux was inferred from the flux excess in stacked IRAC $5.8\,\mu m$ band to IRAC $3.6\,\mu m$ band, and $\xi_{ion}$ was further derived from the UV continuum computed from the stacked SED with the assumption of no dust correction. Their result constituted one of the largest $\xi_{ion}$ estimation.

\textbf{\citet{Castellano23} calculated the $\xi_{ion}$ of more than 1000 VANDELS galaxies at $2.5 < z < 5$ from a multi-band SED fitting approach with BEAGLE. They found no clear evolution of $\xi_{ion}$ with redshift within their probed range, but clear correlations with respect to stellar mass. Also \citet{Saldana-Lopez23} directly measured $\xi_{ion}$ of the LAEs and non-LAEs from their rest-frame UV spectra. Their LAEs have $\sim0.3$ dex higher $\xi_{ion}$ respect to normal non-LAEs.}

\citet{Matthee22} obtained emission line fluxes and physical properties for a sample of 117 O3Es at $z\simeq5.3-6.9$, using the deep \emph{JWST/NIRCam} wide field slitless spectroscopic observations. Measurements of physical properties in their study were based on the median stack spectra of O3Es and following SED fiting. $\xi_{ion}$ was obtained from H$\beta$ emission line and SED derived UV luminosity. Also, a subset of 58 O3Es had spectral coverage of H$\gamma$ with $E(B-V)_{neb}=0.14$ estimated from the observed H$\gamma$/H$\beta$ ratio. 

\citet{Whitler23} selected a candidates of 27 galaxies from the large-scale galaxy overdensities surrounding UV luminous LAEs in the CEERS \emph{JWST/NIRCam} imaging \citep{Bagley23}. They modelled the SEDs of these galaxies and obtained the SED-inferred $\xi_{ion}$ by using the \robotoThin{BEAGLE} code.

\citet{Arteaga24} performed resolved SED modelling on a highly-lensed galaxy at $z=6.072$ from \emph{JWST/NIRCam} imaging. The SED results retrieve consistent measurement of emission lines to the IFU spectra. Based on the resolved SED fitting results, they provided 2D distribution of $\xi_{ion}$ of their target.

\citet{Saxena23} used \emph{JWST/NIRCam} spectroscopic data from the JADES survey \citep{Eisenstein23} and directly measured $\xi_{ion}$ of 17 faint LAEs. Similarly, \citet{Boyett24} used the same dataset and measured $\xi_{ion}$ of 28 [O{\sc iii}] or H$\alpha$ EELGs. The stellar mass of each EELG is fitted by \robotoThin{BEAGLE}.

We perform a fit to all the listed data points, and the results indicate a clear evolution of $\xi_{ion}$ with redshift, which is consistent with previous studies in the literature \citep[e.g.,][]{Shivaei18,Atek22}. The best-fitting relation between $\xi_{ion}$ and redshift yields the following result,
\begin{equation}
    \label{equ:xiionz}
    \log \xi_{ion}= (0.10\pm0.02)\times z + (24.92\pm0.10).
\end{equation}

The intrinsic differences among the sample selection in different studies should be taken into consideration. The current limitations in data availability make it challenging to identify a large number of individual galaxies at higher redshifts. Stacking methods, while useful, can introduce systematic differences among the samples.

Constructing large samples of individual low-mass galaxies is a challenging task, and currently, it has been undertaken in \citet{Atek22} and our study. To obtain a clearer understanding of $\xi_{ion}$ at higher redshifts, it is crucial to construct large samples of individual low-mass galaxies. 

\section{Discussion}
\label{sec:dis}
\subsection{"Downsizing" relation between $\xi_{ion}$ and $M_*$?}
\begin{figure}[hbt!]
    \vspace{0.4cm}
    \includegraphics[width=1\linewidth]{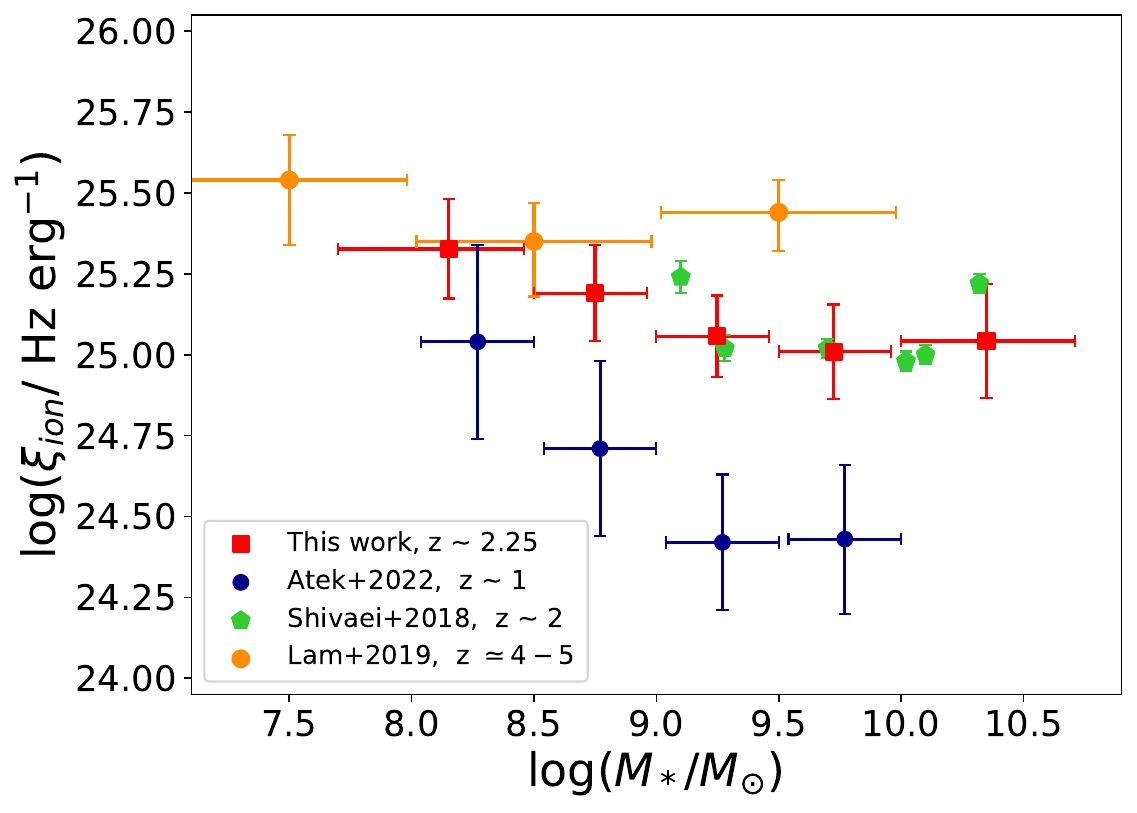}
    \label{fig:xiionmassz}
    \vspace{-0.4cm}
    \caption{The relation between $\xi_{ion}$ and stellar mass at different redshifts. Red squares are stacks in 5 mass bins, same as the left panel in Figure \hyperref[fig:xiion_mass]{9}. A possible downsizing of the relation between $\xi_{ion}$ and $M_{\odot}$ with the increase of redshift is shown here.}
\end{figure}

In section \hyperref[sec:xiionmass]{4.2.1}, we found an enhancement of $\xi_{ion}$ at low-masses. We compared our findings with several studies that also examined $\xi_{ion}$ and $M_*$ relationships below $10^9M_\odot$, but at different redshifts. \citet{Atek22} has analyzed 3D-HST galaxies at $z\sim1$ and also observed an enhancement of $\xi_{ion}$ at lower mass, around 0.5 dex, which is larger than our sample's $\sim0.2$ dex enhancement. Conversely, \citet{Lam19} has stacked IRAC images of galaxies at $z\simeq4-5$ and found an almost independent relationship between $\xi_{ion}$ and $M_*$. By combining our result at $z\sim2.3$ in Figure \hyperref[fig:xiionmassz]{9}, we speculate a possible "downsizing" relationship between $\xi_{ion}$ and $M_*$ over cosmic time, suggesting that the correlation between $\xi_{ion}$ and $M_*$ weakens from lower redshift to higher redshift. 

\begin{table*}[hbt!]
    \small
    \centering
    \label{tab:sps}
    \caption{Various SPS models and corresponding $\xi_{ion}$}
    \begin{tabular}{cccc}
        \hline\hline
        $\ $ Model (upper mass cutoff) $\ $& Metallicity ($Z_*$) & SFH \& stellar age & $\ $$\log(\xi_{ion,model}/\left[\mathrm{Hz}\,\mathrm{erg}{ }^{-1}\right])$$\mathrm{\,^{a}}$ $\ $\\
        \hline
        BPASS$\,$v2.2.1 (300$\,M_{\odot}$) & 0.0001 & Constant, $t=10^8\,$yr & 25.56  \\
        BPASS$\,$v2.2.1 (300$\,M_{\odot}$) & 0.002 & Constant, $t=10^8\,$yr & 25.51  \\
        BPASS$\,$v2.2.1 (100$\,M_{\odot}$) & 0.002 & Constant, $t=10^8\,$yr & 25.40  \\
        BPASS$\,$v2.2.1 (100$\,M_{\odot}$) & 0.004 & Constant, $t=10^8\,$yr & 25.37 \\
        BPASS$\,$v2.2.1 (100$\,M_{\odot}$) & 0.014 ($Z_{\odot}$) & Constant, $t=10^8\,$yr & 25.21  \\
        BPASS$\,$v2.2.1 (100$\,M_{\odot}$) & 0.004 & Constant, $t=10^{7.5}\,$yr & 25.46  \\
        BC03$\,$v2016 (100$\,M_{\odot}$) & 0.004 & Constant, $t=10^8\,$yr & 25.16  \\
        \hline
    \end{tabular}
    \begin{tablenotes}
    \item \textbf{Notes.} ${ }^{\text {a }}$ $\xi_{ion,model}$ is derived from the ionizing photon production rate and the luminosity in the FUV band from the spectral synthesis outputs of the models.
    \end{tablenotes}
\end{table*}

\begin{figure}[hbt!]
    \vspace{0.4cm}
    \includegraphics[width=1\linewidth]{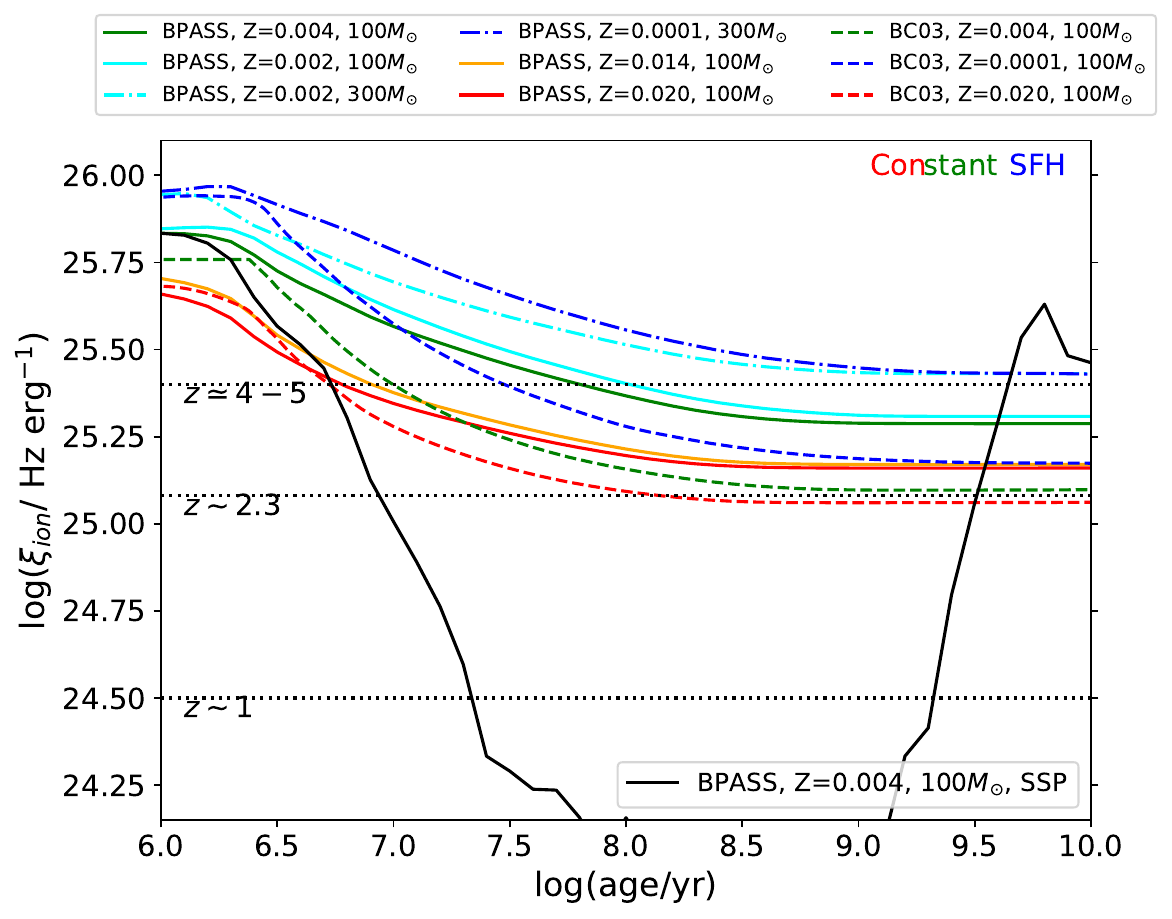}
    \label{fig:xiionmodel}
    \vspace{-0.4cm}
    \caption{The evolution of $\xi_{ion}$ with stellar age in various SPS models with different upper mass cutoff and metallicities. Models with lower metallcities and higher upper mass cutoff are having higher $\xi_{ion}$ more than a factor of 2. Here, if we suppose a constant SFH, $\xi_{ion}$ would gradually drop but finally converge. For reference, we also include one typical BPASS SSP model as black solid line and observed results for massive galaxies at $z\sim1$, $z\sim2.3$ and $z\simeq4-5$ as black dotted line.}
\end{figure}

At the low-mass end, the difference in $\xi_{ion}$ among various redshifts is smaller than $0.5$ dex. This discrepancy 
could potentially be mitigated or even canceled out if we extrapolate the $M_*-\xi_{ion}$ relationship from our study and \citet{Atek22} to lower masses, around $\mathrm{log}(M_*/M_{\odot})\simeq7.5$, as was in \citet{Lam19}. This trend suggests the possibility of an upper limit for $\xi_{ion}$ theoretically, and the low-mass galaxies are gradually approaching the upper limit $\xi_{ion}$ value.

To explore the theoretical limits of $\xi_{ion}$, we investigate several stellar population synthesis (SPS) models. The models we consider include the “Binary Population and Spectral Synthesis” (BPASS v2.2.1) models \citep{Stanway18} and GALAXEV (BC03 v2016) models \citep{Bruzual03} with different metallicities. The initial mass function is using Chabrier IMF \citep{Chabrier03} but with varying upper mass cutoffs. In Table \hyperref[tab:sps]{3} and Figure \hyperref[fig:xiionmodel]{10}, we present the results from these models under the assumption of a constant SFH. If we assume a stellar age of 30$\,$Myr in the constant SFH scenario, the estimation of $\xi_{ion}$ from the BPASS model with an upper mass cutoff of 100$\,M_{\odot}$ and sub-solar metallicity is consistent with the observed $\log \xi_{ion}\sim25.5$ at $z\simeq4-5$. It should be noted that the actual stellar ages may vary among the samples, and even younger stellar populations with an age of 10$\,$Myr would exhibit $\sim0.15\,$dex higher $\xi_{ion}$ compared to those at 30$\,$Myr. This estimation is in closer agreement with recent observations from \emph{JWST} at $z>8$ \citep{Whitler23}.
Also, the realistic SFHs of galaxies are more complex than a simple assumption of a constant SFH. If a galaxy's star formation rate is increasing, it is expected to have a higher $\xi_{ion}$ than that of a galaxy with a constant SFH at the same stellar age. 
Given that low-mass galaxies are more likely to have an increasing star formation rate, their upper limits of $\xi_{ion}$ may be even higher.

At the high-mass end, we find that the $\xi_{ion}$ of massive galaxies at $z\sim1$ is almost one magnitude lower than that at $z\simeq4-5$. We propose that these galaxies may be at different stages of their star formation. Those galaxies at $z\sim1$ may have decreasing star formation rates, resulting in the lower observed $\xi_{ion}$. In Figure \hyperref[fig:xiionmodel]{10}, we also exhibit the evolution of $\xi_{ion}$ for a simple stellar population (SSP) from the BPASS model. If a galaxy undergoes quenching, $\xi_{ion}$ would rapidly decrease to $\log \xi_{ion}\sim24.0$ within $10^8\,$yr. 
On the other hand, massive galaxies at higher redshift in Figure \hyperref[fig:xiionmassz]{9} may still be undergoing continuous star formation. The discrepancy in their $\xi_{ion}$ is likely due to differences in their stellar populations. Referring to Figure \hyperref[fig:xiionmodel]{10}, $\xi_{ion}$ would gradually decrease but eventually converge after $10^9\,$yr. For comparison, we include the observed results for massive galaxies at various redshifts in Figure \hyperref[fig:xiionmodel]{10}. It is suggested that the BPASS models would be more representative for galaxies at $z\simeq4-5$, while the BC03 models for our sample at $z\sim2.3$.

\subsection{Implications for Reionization}
Frameworks on cosmic reionization \citep[e.g.,][]{Robertson13, Robertson15} highlighted the importance of constraining two key parameters: the escape fraction $f_{esc}$ and the ionizing photon production efficiency $\xi_{ion}$, at $z>6$. \citet{Bouwens15} has established a well-matched evolution of the cosmic ionizing emissivity and the galaxy UV-luminosity density, suggesting that galaxies are the major sources responsible for reionization. They also constrained a lower limit of $\log (\xi_{ion}\,f_{esc})=24.50\pm0.1$, if only galaxies contributed to the reionization.

In our study, we assume that the low-mass HAEs at $z\sim2.3$ can serve as the analog population of the galaxies that reionized the universe at $z>6$. The median $\xi_{ion}$ of these low-mass HAEs is comparable to the canonical value of $\log \xi_{ion}=25.2$ with an escape fraction $f_{esc}=0.2$ \citep{Robertson13,Bouwens15}. Note that our measurement of $\xi_{ion}$ assumes no escape fraction. Therefore, we can infer that the required escape fraction to ionize the universe at $z>6$ is likely no larger than 0.2.

However, the assumption that low-mass galaxies at $z\sim2$ have similar physical properties to those at higher redshift may not be entirely accurate. For instance, low-mass galaxies at higher redshift are likely to exhibit higher $\xi_{ion}$ compared to our sample.
Hydrodynamical simulations show that successive starburst activities are prevalent in low-mass galaxies \citep[e.g.,][]{Dominguez15,Sparre2017,Emami19}. While, variations in the underlying physical conditions of the interstellar medium (ISM) over cosmic time, such as changing metallicities, can lead to different levels of starbursts, resulting in varying equivalent widths of emission lines and different observed $\xi_{ion}$ at different redshifts.
In section \hyperref[sec:xiionz]{5.1}, we inferred a possible upper limit of $\xi_{ion}$ at $\log \xi_{ion}\gtrsim25.5$. If we accept this value and assume $\log (\xi_{ion}\,f_{esc})=24.60$, $f_{esc}$ should be 0.13 at maximum for low-mass galaxies during reionization to reionize the universe.


Observations of LyC signal and the corresponding $f_{esc}$ at $z\simeq2-4$ have been conducted by several studies in the past decade. Some surveys have reported very few individual detection of LyC emission, but inferred esacpe fraction $f_{esc} \sim 0.05-0.1$ based on stacking imaging \citep[e.g.,][]{Marchi17, Naidu18,Steidel18,Pahl21}. Still, there are also several individual galaxies that have been observed with very high escape fraction \citep[e.g.,][]{MarquesChaves22}.
The above results, comparable to our estimation, indicate that galaxies could be the dominant source responsible for reionizing the universe.

\section{CONCLUSIONS}
\label{sec:conclu}
In this work, we have investigated the galaxy properties, including SFR, $\xi_{ion}$, UV properties, equivalent width, of the 401 low-mass ($<10^9\,M_{\odot}$) HAEs from C24 in three ZFOURGE fields. The selection of HAEs is based on the flux excess in ZFOURGE-$K_s$ filters due to the intense $\mathrm{H\alpha}$ emission lines relative to the best-fit stellar continuum from SED fitting. The main results can be summarized as follows:
\begin{itemize}[leftmargin=*]
    \item The $\mathrm{SFR}-M_*$ relation, i.e., SFMS, derived from $\mathrm{H\alpha}$ and UV luminosity exhibits a similar slope ($0.56\pm0.03$ vs. $0.60\pm0.04$) above the stellar mass completeness. The low-mass HAEs ($<10^9\ M_{\odot}$) scatter above the SFMS($\mathrm{H\alpha}$) by $\sim$0.3 dex, while this scattering is not evident in the SFMS(UV).
    
    \item Analysis of the $\mathrm{SFR(H\alpha)}$/$\mathrm{SFR(UV)}$ ratio and star formation history suggests that these low-mass HAEs are undergoing early-stage bursty star formation phases with rising SFR. This is reflected in the abundance of younger stellar population with high $EW_{\mathrm{H\alpha}}\sim$1000 $\mathrm{\AA}$ in the galaxies. Our population may be the analogs of the major contributors to cosmic reionization at $z>6$.
    
    \item The ionizing photon production efﬁciency, $\xi_{ion}$, of the low-mass HAEs is found to be $\mathrm{log}(\xi_{ion}/erg^{-1} Hz)=25.24^{+0.10}_{-0.13}\  (25.35^{+0.12}_{-0.15})$, assuming the Calzetti (SMC) curve for UV dust correction. This result is higher by $\sim$0.2 dex compared to other HAEs in our study, suggesting a possible mass dependence of $\xi_{ion}$.
    
    \item We observe a correlation between $\xi_{ion}$ and both the UV slope ($\beta_{\mathrm{UV}}$) and UV absolute magnitude ($M_{\mathrm{UV}}$). Galaxies with bluest UV slopes and faintest UV luminosities exhibit an enhanced $\xi_{ion}$ by nearly a factor of 2 compared to the median $\xi_{ion}$ of our sample.

    \item Our results confirm that $\xi_{ion}$ is related to the equivalent widths of H$\alpha$ and [O{\sc iii}] ($EW_\mathrm{{[O\pnt{III}]}}$ and $EW_\mathrm{{H\alpha}}$). This indicates that the equivalent width of these strong optical lines can serve as a proxy for estimating $\xi_{ion}$. However, the relationship appears to be less steep compared to literature results, suggesting differences in the galaxy properties of the studied samples.
\end{itemize}

By combining a comprehensive analysis of literature results, we have strengthened the evidence for the evolution of $\xi_{ion}$ with redshift, extending our study to a significant number of low-mass galaxies at $z\sim2$. Moreover, our findings suggest a potential "downsizing" relationship between $\xi_{ion}$ and stellar mass as we trace back in cosmic time. We utilize stellar population synthesis (SPS) models to highlight that $\xi_{ion}$ observed in low-mass galaxies may approach the upper limit predicted by these models. This finding suggests that low-mass galaxies could be reaching the maximum efficiency of ionizing photon production according to current SPS models. This insight further supports the notion that low-mass galaxies play a crucial role in cosmic reionization.

To further enhance our understanding, the vast amount of data from the James Webb Space Telescope (JWST) will play a crucial role. This will enable the construction of a large sample of low-mass HAEs at higher redshifts, including even lower mass regimes, allowing for a more comprehensive exploration of cosmic reionization.

\vspace{5mm}

The authors thank referee for the helpful suggestions and comments. We are grateful for enlightening conversations with Jonathan Cohn, Ken-ichi Tadaki, and Kazuki Daikuhara. We gratefully thank Hakim Atek for the great results from the 3D-HST dataset. We also gratefully thank Mengtao Tang and Najmeh Emami for the best-fit relation results in their previous work. Data analysis was in part carried out on the Multi-wavelength Data Analysis System operated by the Astronomy Data Center (ADC), National Astronomical Observatory of Japan. 
This work was supported by the Graduate Research Abroad in Science Program (GRASP) from the University of Tokyo and MEXT/JSPS KAKENHI grant Nos. 15H02062, 24244015, 18H03717, 20H00171, 22K21349.

\vspace{5mm}

\section*{\textbf{ORCID iDs}}
\noindent
Nuo Chen \orcidlink{0000-0002-0486-5242}\url{https://orcid.org/0000-0002-0486-5242}\\
Kentaro Motohara \orcidlink{0000-0002-0724-9146}\url{https://orcid.org/0000-0002-0724-9146}\\
Lee Spitler \orcidlink{0000-0001-5185-9876}\url{https://orcid.org/0000-0001-5185-9876}\\
Kimihiko Nakajima \orcidlink{0000-0003-2965-5070}\url{https://orcid.org/0000-0003-2965-5070}

\bibliography{main}{}
\bibliographystyle{aasjournal}

\end{document}